# Physical properties of new MAX phase borides M₂SB (M = Zr, Hf and Nb) in comparison with conventional MAX phase carbides M₂SC (M = Zr, Hf and Nb): Comprehensive insights

M. A. Ali[a,*], M. M. Hossain[a], M. M. Uddin[a], M. A. Hossain[b], A. K. M. A. Islam[c,d], S. H. Naqib[c,*]

[a]Department of Physics, Chittagong University of Engineering and Technology (CUET), Chattogram4349, Bangladesh

[b]Department of Physics, MawlanaBhashani Science and Technology University, Santosh, Tangail-1902, Bangladesh

[c]Department of Physics, University of Rajshahi, Rajshahi-6205, Bangladesh

[d]Department of Electrical and Electronic Engineering, International Islamic University Chittagong, Kumira, Chittagong, 4318, Bangladesh

## Abstract

In this article, a detailed study of the recently synthesized MAX phase borides M₂SB (M = Zr, Hf and Nb) has been performed via first principles technique. Investigation of mechanical hardness, elastic anisotropy, optical properties, dynamical stability and thermal properties are considered for the first time. The estimated values of stiffness constants and elastic moduli are found in good agreement with available results. The micro and macro hardness ($H_{\mathrm{micro}}$ and $H_{\mathrm{macro}}$) parameters are calculated. The Vickers hardness is also calculated using Mulliken population analysis. The electronic density of states and charge density mapping are used to explain the variation of stiffness constants, elastic moduli and hardness parameters among the studied ternary borides. The Nb₂SB compound is found to show best combination of mechanical properties. Mixture of covalent and ionic bonding within these borides is explained using Mulliken population analysis. The direction dependent values of Young's modulus, compressibility, shear modulus and Poisson's ratio are visualized by 2D and 3D representations and different anisotropic factors are calculated. The important optical constants are calculated and analyzed. The metallic nature of the studied borides is confirmed from the DOS and optical properties. The reflectivity spectra reveal the potential use of Zr₂SB as coating materials to diminish solar heating. The studied borides are dynamically stable as confirmed from the phonon dispersion curves. The characteristic thermodynamic properties have also been calculated and analyzed. The physical properties of corresponding 211 MAX phase carbides are also calculated for comparison with those of the titled ternary borides.

## 1. Introduction

MAX phase is one of the exciting families of materials, generally expressed as $M_{n+1}AX_n$, n = 1, 2, 3, where early transition metals are M; an element belonging to groups 12-16 is A and X is usually either C or N, that was first brought to front in 1960s by Nowotny et al [1-4]. Barsoum et al. [5,6] have revived the interest in the 1990s by disclosing their outstanding properties. They possess high electrical and thermal conductivity, machinability as well as mechanical strength like metals associated with the excellent mechanical properties at high-temperature, very good corrosion and oxidation resistance like ceramics [7]. The chemistry of this amazing combination of properties lies in the nanolaminated layered structure of MAX phase materials in which a single atomic layer of A (A = Al, Si, Ge, etc.) element is sandwiched in between the $M_{n+1}X_n$ sheets [8]. Owing to the attractive combination of properties suitable for diverse fields of application, the study of viable MAX phase materials has been recognized as a vital sub-field of materials science research [9]. Consequently, the published articles on the MAX phases have found to have an exponential increase in recent years, reflecting the interest and quick improvement in the field [10].

*Corresponding authors: ashrafphy31@cuet.ac.bd; salehnaqib@yahoo.com

The prospective demands of the MAX phase materials are motivating researchers to improve the diversity and performance by reporting new compounds belonging to the MAX phase or by mixing of the M, A and/or X elements of the existing phases [11-26]. The studies regarding the effects of point defects and incorporation of atoms have also been reported [27–30]. However, most of the diversity is limited to the M- and A- elements and until recent times the X element is kept as either C and/or N [31]. Recently, the diversity has been expanded by incorporating B (Boron) as an X element. Researchers are quite hopeful with the MAX phase borides because of the interesting physical and chemical properties of B and its compounds [32]. Hence, along with the conventional MAX phase materials, MAX phase borides are also expected to become promising candidates from application as well as research point of view. Consequently, some reports on the MAX phase borides have already been published [33-39]. Khazaei et al. [33] have investigated the formation energies, electronic and mechanical properties of hypothetical $M_2AlB$ (M = Sc, Ti, Cr, Zr, Nb, Mo, Hf, or Ta) MAX phase borides. Genceret al. [34] have studied the electronic and lattice dynamical properties of $Ti_2SiB$. A group of hypothetical MAX phase boride compounds [$M_2AB$ (M = Ti, Zr, Hf; A = Al, Ga, In)] have been investigated by Surucu et al. [35]. Chakraborty et al. [10] have predicted the soft MAX ($V_2AlB$) phase by Boron substitution. Miao et al [36] have predicted some Boron based MAX phases which are thermodynamically stable. Rackl et al. [37] have synthesized the Nb based 211 MAX phase boride and their solid solutions where C is substituted by B in the $Nb_2SC$. The lattice dynamical and thermo-elastic properties of $M_2AlB$ (M = V, Nb, Ta) MAX phase borides have been investigated by Surucu et al. [38]. Very recently, the MAX phase borides $Zr_2SB$ and $Hf_2SB$ have been synthesized by Rackl et al. [39]. However, the theoretical studies on $M_2SB$ (M = Zr, Hf and Nb) have been limited to electronic and elastic properties only. Some other decisive properties should be disclosed that may extend their field of applications. First of all, study of dynamical stability is important concerning application point of view. The knowledge of mechanical anisotropy is crucial for structural materials because it is closely related with some important mechanism such as anisotropic plastic deformation, crack formation and propagation, and elastic instability that limits their usefulness. The Vickers hardness measures the overall bonding strength attributed from the individual bonds of a solid. Since the MAX phase carbides are popular for structural design, MAX phase borides are also expected to exhibit their suitability as structural materials; the Vickers hardness is an important parameter in this regard. Understanding of thermodynamic properties is required to predict the suitability for high temperature and high pressure applications. Conventional MAX phase materials have useful optical characteristics. The MAX phase borides are therefore, expected to exhibit their suitability as materials that can be used in optical device applications. Hence the optoelectronic properties of $M_2SB$ (M = Zr, Hf and Nb) need to be explored. These are the prime motivations of this particular study.

Therefore, a first time study of the mechanical anisotropy, Vickers hardness, thermodynamics and optical properties of $M_2SB$ (M = Zr, Hf and Nb) along with a revisit of electronic and elastic properties will be presented in this paper. It should be noted that we have also calculated the properties of titled carbides ($M_2SC$) for a comprehensive comparison. This will provide us with important insights regarding the role of boron atom in place carbon in ternary 211 MAX phase nanolaminates.

## 2. Calculation methods

The results of $M_2SB$ (M = Zr, Hf and Nb) MAX phase borides have been explored by the density functional theory (DFT) [40, 41] as implemented in the CAmbridge Serial Total Energy Package (CASTEP) code [42]. The first-principles calculations are performed by the plane-wave pseudopotential method. The functional used for the exchange and correlations terms is the generalized gradient approximation (GGA) of the Perdew-Wang 91 version (PW91) [43]. The pseudo-atomic calculations were performed for B - $2s^2\,2p^1$, S - $3s^2\,3p^4$, Zr- $4d^2\,5s^2$, Nb - $4d^4$ $5s^1$ and Hf - $5d^2\,6s^2$ electronic orbitals. The convergence was ensured by setting the cutoff energy at 500 eV and a k-point mesh of size 10×10×3 [44]. Density mixing was used to the electronic structure and Broyden Fletcher Goldfarb Shanno (BFGS) geometry optimization [45] was applied to optimize the atomic configurations. The self-consistent convergence of the total energy is $5 \times 10^{-6}$ eV/atom, and the maximum force on the atom is 0.01 eV/Å. The maximum ionic displacement is set to $5 \times 10^{-4}$ Å, and a maximum stress of 0.02 GPa is used.

## 3. Results and discussion

### 3.1. *Structural properties*

Fig. 1 shows the unit cell of the $Zr_2SB$ boride which is found to be crystallized with hexagonal structure [Space group$P63/mmc$] like the conventional C/N containing MAX phases [7]. The unit cell also consists of two formula units and there are four atoms per formula unit cell. The atomic positions of Zr, S and B atoms in the unit cell are $(1/3, 2/3, z_M)$,$(1/3, 2/3, 3/4)$ and $(0, 0, 0)$, respectively. The $z_M$ is known as internal parameter; the values of $z_M$ are given in Table 1.

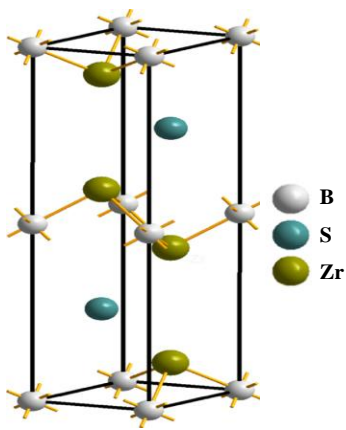

**Fig. 1 - Crystal structure of the $Zr_2SB$ compound.**

The optimized lattice constants ($a$, $c$), internal parameters and $c/a$ ratios of $M_2SB$ (M = Zr, f and Nb) are listed in Table 1 along with the reported results and their corresponding carbides. It is observed from Table 1 that the calculated values of $a$ and $c$ are very close to the experimental values. The maximum deviation (1.13 %) from the experimental lattice constant is observed for $a$ of $Hf_2SB$ phase. A very good agreement is also observed between theoretically obtained values (obtained in this work and by Rackl et al [37, 39]). This confirms a very high level of accuracy of the present study.

**Table 1 - Calculated lattice parameters (*a* and *c*), internal parameter ($z_M$) and *c/a* ratio of M₂SX (M = Zr, Hf and Nb; X = B and C) MAX phases.**

| Phase | *a* (Å) | % of deviation | *c* (Å) | % of deviation | $z_M$ | *c/a* | Reference |
|-------|---------|----------------|---------|----------------|-------|-------|-----------|
| Zr₂SB | 3.5160 | 0.45 | 12.3137 | 0.34 | 0.6052 | 3.502 | This |
|       | 3.5001 |      | 12.2712 |      | 0.6060 | 3.505 | Expt. [39] |
|       | 3.519  | 0.08 | 12.317  | 0.02 | 0.6055 | 3.500 | Theo. [39] |
| Zr₂SC | 3.4201 | 0.24 | 12.2053 | 0.49 | 0.6007 | 3.568 | This |
|       | 3.4117 |      | 12.1452 |      | 0.6013 | 3.560 | Expt. [39] |
|       | 3.423  | 0.08 | 12.226  | 0.17 | 0.6006 | 3.572 | Theo. [39] |
| Hf₂SB | 3.5064 | 1.13 | 12.1592 | 0.45 | 0.6039 | 3.468 | This |
|       | 3.4671 |      | 12.1046 |      | 0.6047 | 3.491 | Expt. [39] |
|       | 3.482  | 0.70 | 12.137  | 0.18 | 0.6038 | 3.485 | Theo. [39] |
| Hf₂SC | 3.4247 | 1.63 | 12.1847 | 1.39 | 0.6005 | 3.558 | This |
|       | 3.3695 |      | 12.0172 |      | 0.6004 | 3.566 | Expt. [39] |
| Nb₂SB | 3.3583 | 0.69 | 11.6043 | 0.47 | 0.6009 | 3.455 | This |
|       | 3.335  |      | 11.55   |      | 0.6017 | 3.463 | Expt. [37] |
| Nb₂SC | 3.2963 | 0.56 | 11.6699 | 1.56 |        | 3.540 | This |
|       | 3.278  |      | 11.49   |      |        | 3.505 | Expt. [37] |

### 3.2 Mechanical properties

Although, Rackl et al. [37, 39] have studied the elastic properties of the titled borides but we have calculated these again with the intention: (i) to assess the present results by comparing with their results and (ii) to calculate some unexplored parameters such as machinability index, Cauchy pressure and hardness parameters etc.

*The stiffness constants: mechanical stability, machinability index and Cauchy pressure*

The stiffness constants ($C_{ij}$) are the basis to evaluate the mechanical performance of materials. Consequently, their study is essential regarding practical applications in various technologies. Moreover, these constants can also be used to unveil the mechanical stability, polycrystalline elastic moduli, ductile/brittle nature, machinability index, anisotropic behavior, hardness etc. Thus, we have calculated the stiffness constants ($C_{ij}$) of the synthesized borides [M₂SB (M = Zr, Hf and Nb)] as well as conventional MAX phase carbides [M₂SC (M = Zr, Hf and Nb)] as listed in Table 2 along with other available results for comparison. We would like to start our discussion by checking the mechanical stability of studied borides and carbides. The examination of mechanical stability is required to evaluate material's sustainability in any application where static stress is applied. The conditions for mechanical stability of a hexagonal system are: $C_{11} > 0$, $C_{33} > 0$, $C_{11}-C_{12} > 0$, $C_{44} > 0$, $(C_{11} +C_{12})C_{33}- 2(C_{13})^2 > 0$ [48]. The values listed in Table 2 satisfy these inequalities relations, suggesting the mechanically stability of the MAX compounds under investigation. In addition, some other information can also be extracted from the obtained stiffness constants. For instance, the $C_{11}$ and $C_{33}$ correspond to the stiffness for applied stress along [100] and [001] directions, respectively, and $C_{44}$ assess the resistance against shear deformation in the (100) plane. The $C_{11}$ and $C_{33}$ are larger than $C_{44}$ revealing a low resistance required for shear deformation compared to unidirectional deformation for borides considered here. Moreover, the $C_{11} < C_{33}$ revealing a low resistance to deformation along crystallographic *a* - axis than that along $c$ – axis. It is also observed that the $C_{11}$ is lower than $C_{33}$ of the M$_2$SC (M = Zr, Hf and Nb) carbides. The anisotropic bonding strength along *a* and *c*-axis is also revealed by the unequal values of $C_{11}$ and $C_{33}$. The values of $C_{12}$ and $C_{13}$ are comparatively smaller than other single crystal elastic constants. These two factors combine an applied stress component along the *a*-axis with a uniaxial strain in the *b*- and *c*-axis, respectively. Furthermore, the obtained stiffness constants of M$_2$SX (M = Zr, Hf and Nb; X = B and C) phases are displayed in Fig. 2 (a) for better visualization of the variation among them.

The machinability index, a useful performance indicator also be predicted using $C_{44}$ by the formula $B/C_{44}$ [49]. The $B/C_{44}$ ratio is observed to increase for M$_2$SB (M = Zr, Hf) compounds compared to the M$_2$SC (M = Zr, Hf) carbides and decreases for the Nb$_2$SB compared to Nb$_2$SC. In this case, the value of $C_{44}$ (also $C_{11}$ and $C_{33}$) of Nb$_2$SB is higher than Nb$_2$SC but the value of $B$ (Nb$_2$SB) is still lower than that of Nb$_2$SC, leading to a higher $B/C_{44}$ ratio. It is found that the phase having low shear resistance (small $C_{44}$) exhibits high machinability. It is noted that the W$_2$SnC has shown highest machinability (33.33) with the lowest $C_{44}$ (6 GPa) among all the 211 MAX phases [50].

**Table 2 - Calculated stiffness constants ($C_{ij}$), bulk modulus ($B$), machinability index ($B/C_{44}$), Cauchy pressure ($CP$), shear modulus ($G$), Young's modulus ($Y$), Poisson's ratio ($v$), Pugh ratio ($G/B$), micro hardness ($H_{micro}$) and macro hardness ($H_{macro}$) of $M_2SX$ (M = Zr, Hf and Nb; X = B and C) MAX phases.**

| Parameters | $Zr_2SB$ | $Zr_2SC$ | $Hf_2SB$ | $Hf_2SC$ | $Nb_2SB$ | $Nb_2SC$ |
|---|---|---|---|---|---|---|
| $C_{11}$ (GPa) | 256 | 295 | 285 | 311 | 326 | 316 |
| | 261[a] | 296[b] | 286[a] | 344[c] | 316[d] | 301[d] |
| $C_{33}$ (GPa) | 280 | 315 | 298 | 327 | 328 | 325 |
| | 282[a] | 316[b] | 296[a] | 369[c] | 317[d] | 314[d] |
| $C_{44}$ (GPa) | 115 | 138 | 130 | 149 | 151 | 124 |
| | 117[a] | 140[b] | 122[a] | 175[c] | 143[d] | 116[d] |
| $C_{12}$ (GPa) | 85 | 89 | 81 | 97 | 86 | 108 |
| | 79[a] | 90[b] | 79[a] | 116[c] | 95[d] | 105[d] |
| $C_{13}$ (GPa) | 79 | 102 | 92 | 121 | 126 | 151 |
| | 80[a] | 102[b] | 84[a] | 138[c] | 131[d] | 157[d] |
| $B$ (GPa) | 142 | 166 | 156 | 181 | 186 | 197 |
| | 142[a] | 166[b] | 151[a] | 204[c] | 186[d] | 194[d] |
| $B/C_{44}$ | 1.23 | 1.20 | 1.20 | 1.21 | 1.23 | 1.58 |
| | 1.21[*] | 1.18[*] | 1.23[*] | 1.16[*] | 1.30[*] | 1.67[*] |
| $CP$ (GPa) | -30 | -49 | -49 | -52 | -65 | -16 |
| $G$ (GPa) | 99 | 115 | 112 | 120 | 128 | 105 |
| | 102[a] | 116[b] | 111[a] | 134[c] | 116[d] | 97[d] |
| $Y$ (GPa) | 241 | 280 | 270 | 295 | 312 | 267 |
| | 247[a] | 282[b] | 267[a] | 330[c] | 287[d] | 249[d] |
| $v$ | 0.22 | 0.22 | 0.21 | 0.23 | 0.22 | 0.27 |
| | 0.21[a] | 0.22[b] | 0.20[a] | 0.23[c] | 0.24[d] | 0.28[d] |
| $G/B$ | 0.70 | 0.69 | 0.71 | 0.66 | 0.68 | 0.53 |
| | 0.71[a] | 0.70[b] | 0.73[a] | 0.65[c] | 0.62[d] | 0.50[d] |
| $H_{micro}$(GPa) | 18.67 | 21.57 | 21.63 | 23.88 | 19.10 | 15.84 |
| | 19.73[*] | 22.52[*] | 22.25[*] | 24.14[*] | 20.05[*] | 14.26[*] |
| $H_{macro}$(GPa) | 16.37 | 17.79 | 18.17 | 17.24 | 18.76 | 11.58 |
| | 17.04[*] | 18.25[*] | 18.75[*] | 18.20[*] | 15.57[*] | 09.91[*] |

[a]Ref-[39], [b]Ref-[46], [c]Ref-[47], [c]Ref-[13]; [*]calculated from literature values.

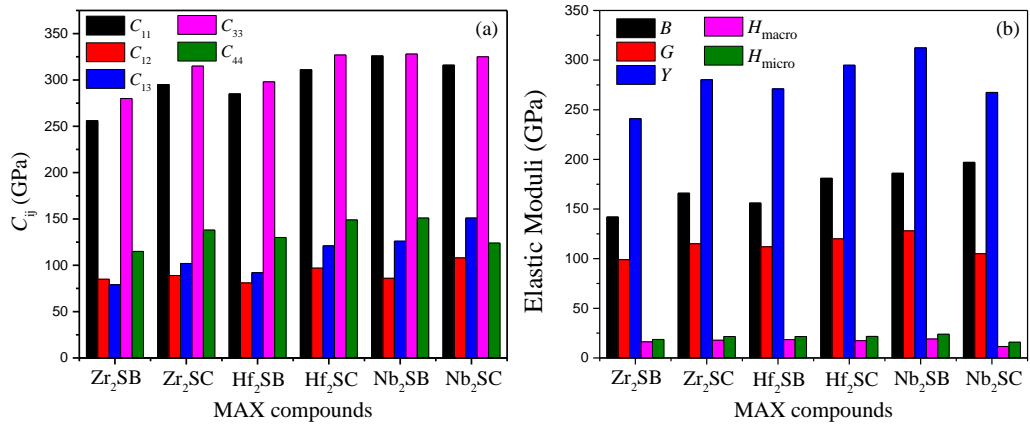

**Fig. 2 - (a) The stiffness constant and (b) elastic moduli of $M_2SX$ (M = Zr, Hf and Nb; X = B and C) MAX phases.**

Moreover, the obtained $C_{ij}$ can also be used to mark out the dominating bonding nature in these borides by calculating the Cauchy pressure (*CP*). The *CP* is defined as the difference between $C_{12}$ and $C_{44}$ that renders us with some information of solids important for practical applications [51]. For example, the type of atomic bonding within the solids can be known from the sign of *CP*; the negative (positive) sign represents the directional covalent (ionic) bonding. It (*CP*) is also an indicator of brittle or ductile nature of solids; the positive (negative) values of *CP* suggest ductile (brittle) nature of solids. The listed values of *CP* (negative values) presented in Table 2 indicate that the borides have predominantly directional covalent bonding and they behave as brittle materials like the conventional MAX phase carbides. Finally, the ranking of the stiffness constants in the studied borides M$_2$SB is as follows: $C_{ij}$(Nb$_2$SB) > $C_{ij}$(Hf$_2$SB) > $C_{ij}$(Zr$_2$SB).

*Elastic moduli and hardness*

We have also calculated the mechanical properties such as the polycrystalline elastic moduli (*B*, *G*, *Y*), brittle or ductile indicator and hardness parameters of the studied borides/carbides from the elastic constants. The well-known Voigt–Reuss–Hill (VRH) approximation [52, 53] is used to calculate the elastic parameters of the polycrystalline borides. The equations used to calculate the Young's modulus (Y) and Poisson's ratio (υ) can be found elsewhere [54, 55]. Observing Table 2, we conclude that our results are in good accord with the earlier reported results and the elastic moduli are smaller for the Zr$_2$SB and Hf$_2$SB borides than the Zr$_2$SC and Hf$_2$SC carbides but elastic moduli of Nb$_2$SB are larger than that of Nb$_2$SC except the bulk modulus; the *B* of Nb$_2$SB is smaller than that of Nb$_2$SC as shown in Fig. 2(b) and Table 2. It is noted that the shear stiffness constant $C_{11}$, $C_{33}$ and $C_{44}$ are also higher for the same compound. These parameters are not related directly to hardness but are normally larger for harder materials [56]. Another point is noted here that the brittleness indicators (Poisson's and Pugh ratio [57, 58]) predicted more brittle behavior of the M$_2$SB (M = Zr, Hf and Nb) borides than the M$_2$SC (M = Zr, Hf and Nb) carbides.

The important information of the physical processes of solids such as plastic deformation, depth of indentation, penetration and scratching can be known from the hardness of the solids. Study of hardness is fundamental in the field of engineering for designing devices. The hardness value obtained by experiments depends on the measurement techniques. The theoretically obtained values of hardness also depend on the formalism used for its calculation. We have calculated hardness values of the studied borides/carbides using three different formalisms. The calculated elastic moduli are used to calculate the hardness parameters ($H_{micro}$ and $H_{macro}$) of studied borides/carbides by the following relations: $H_{micro} = \frac{(1-2\nu)E}{6(1+\nu)}$ [54] and $H_{macro} = 2[(\frac{G}{B})^2 G]^{0.585} - 3$ [59]. Although, these hardness parameters ($H_{micro}$ and $H_{macro}$) cannot predict the exact hardness of the materials but can be used to obtain a tentative explanation of variation of the hardness among the solids belonging to a same family crystallized in the same structure. The calculated values of $H_{macro}$ of $M_2SB$ (M = Zr, Hf and Nb) MAX borides are higher than that of $M_2AB$ (M = Ti, Zr and Hf; A = Al, Ga and In) borides [35]. We have also calculated another hardness parameter, the Vickers hardness which is the geometrical averages of hardness of all the bonds present in the solid. Gao [60] proposed a method to calculate the Vickers hardness based on Mulliken population for non-metallic covalent bond dominated materials. Later, Gou et al. [61] corrected the formula for partially metallic bonded compounds such as the studied borides and carbides, which has been widely used in recent times [62-64]. The Mulliken bond populations are used to obtain Vickers hardness of the $M_2SX$ (M = Zr, Hf and Nb; X = B and C) MAX phases. The relevant formula for the hardness is given as [65]:

$$H_V = \left[ \prod^{\mu} \left\{ 740 \left( P^{\mu} - P^{\mu'} \right) \left( v_b^{\mu} \right)^{-5/3} \right\}^{n^{\mu}} \right]^{1/\sum n^{\mu}},$$

where $P^{\mu}$ is the Mulliken population of the $\mu$-type bond, $P^{\mu'} = n_{free}/V$ is the metallic population, and $v_b^{\mu}$ is the bond volume of $\mu$-type bond.

**Table 3 - Calculated Mulliken bond number $n^{\mu}$, bond length $d^{\mu}$, bond overlap population $P^{\mu}$, metallic population $P^{\mu'}$, bond volume $v_b^{\mu}$, bond hardness $H_v^{\mu}$ of $\mu$-type bond and Vickers hardness $H_v$ of $M_2SB$ (M = Zr, Hf and Nb) along with the results of $M_2SC$ (M = Zr, Hf and Nb).**

| Compounds | Bond | $n^{\mu}$ | $d^{\mu}$ (Å) | $P^{\mu}$ | $P^{\mu'}$ | $v_b^{\mu}$ (Å³) | $H_v^{\mu}$ (GPa) | $H_v$(GPa) |
|---|---|---|---|---|---|---|---|---|
| Zr$_2$SB | B-Zr | 4 | 2.40822 | 1.22 | 0.0152 | 18.219 | 7.07 | |
| | S-B | 4 | 2.70175 | 0.76 | 0.0152 | 25.726 | 2.45 | 4.16 |
| Zr$_2$SC | C-Zr | 4 | 2.32604 | 1.07 | 0.0114 | 12.166 | 7.53 | |
| | S-C | 4 | 2.68671 | 0.73 | 0.0114 | 18.734 | 2.48 | 4.33 |
| Hf$_2$SB | B-Hf | 4 | 2.3902 | 1.44 | 0.0050 | 17.778 | 8.76 | |
| | S-B | 4 | 2.70004 | 0.81 | 0.0050 | 25.626 | 2.67 | 4.84 |
| Hf$_2$SC | C-Hf | 4 | 2.32568 | 1.28 | 0.0026 | 16.210 | 9.10 | |
| | S-C | 4 | 2.68858 | 0.80 | 0.0026 | 25.045 | 2.75 | 5.00 |
| Nb$_2$SB | B-Nb | 4 | 2.26543 | 1.22 | 0.0053 | 15.060 | 9.78 | |
| | S-B | 4 | 2.59817 | 0.79 | 0.0053 | 22.721 | 3.18 | 5.58 |
| Nb$_2$SC | C-Nb | 4 | 2.20621 | 0.99 | 0.0155 | 13.679 | 9.21 | |
| | S-C | 4 | 2.62055 | 0.73 | 0.0155 | 22.925 | 2.86 | 5.13 |

The obtained hardness values are listed in Table 2 and Table 3 where the values also exhibit a reflection of the elastic moduli. Like other mechanical properties, the Vickers hardness of $M_2SB$ (M = Zr and Hf) borides is also lower than the carbide $M_2SC$ (M = Zr and Hf) counterparts. The Vickers hardness of $Nb_2SB$, on the other hand, is higher than that of $Nb_2SC$. It is noted that the obtained values of $H_v$ (4.16 to 5.58) GPa are within the general range for MAX phases (2 to 8) GPa [21], indicating their softness and easy machinability similar to other MAX compounds. The lowest value of $H_v$ (4.16 Ga) is obtained for the $Zr_2SB$ which is still higher than some of carbides e.g., $Sc_2InC$ ($H_v \sim 2.4$ GPa) [63] and $M_2InC$ (M = Zr, Hf and Ta; $H_v \sim 1.05$ to 4.12 GPa) [66]. Another point can be noted here that the bond hardness of M-B/C bonds is higher than that of S-B/C bonds, which is another important feature of the layered structure of MAX phase materials [63, 66]. Finally, based on the values of $B$, $G$, $Y$, $H_{micro}$, $H_{macro}$ and $H_v$, the $M_2SB$ (M = Zr, Hf and Nb) borides can be ranked as follows: $Nb_2SB > Hf_2SB > Zr_2SB$. In fact, $Nb_2SB$ exhibits the best combination of mechanical properties among the studied borides and carbides. Now, the question arises what is the physical process behind the variation of the mechanical properties among the studied borides? In order to address this important question, we have calculated the density of states (DOS) and charge density mapping (CDM) of the studied borides $M_2SB$ (M = Zr, Hf and Nb). It is noted that the details of electronic band structure, DOS, electron localization function etc. of the $M_2SB$ (M = Zr, Hf and Nb) borides have been studied by Rackl et al. [37, 39] and we have also calculated the electronic band structure, total and partial DOS of $M_2SB$ (M = Zr, Hf and Nb). We have avoided lengthy discussion on the electronic band structure because details can be found in Refs. [37, 39].

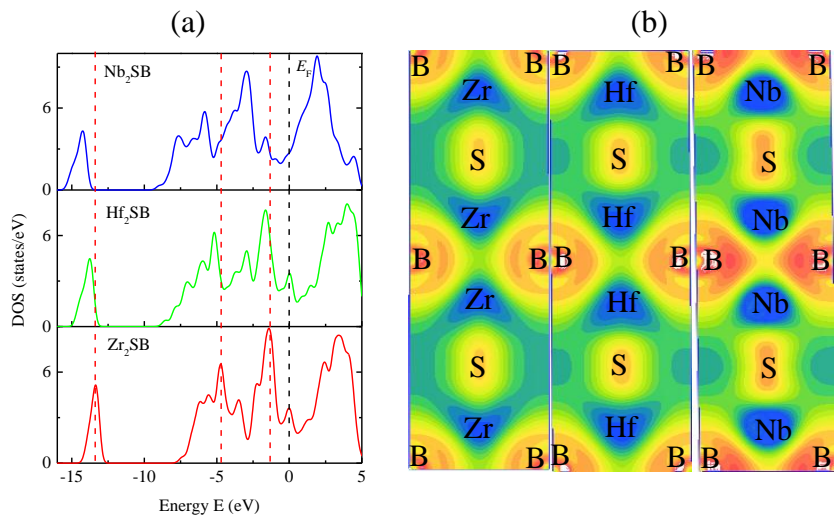

**Fig. 3 - (a) Electronic density of states and (b) charge density mapping of $M_2SB$ (M = Zr, Hf and Nb) borides.**

Fig. 3 (a) shows the total DOS for M$_2$SB (M = Zr, Hf and Nb) borides that exhibit the metallic nature with non-zero finite values of DOS at the Fermi level ($E_\text{F}$). There are three red vertical lines (Fig. 3 (a)) used to demonstrate the positions of the peaks arising from the hybridization between different electronic states for Zr$_2$SB. It is very clear from the reference lines that the peaks are shifted to lower energy when M moves from Zr to Nb. It is found that the hybridized state in the lower energy leads to stronger bonding among the states involved [23, 67]. Thus, the DOS peaks of Nb$_2$SB indicate stronger bonding between the atomic states within this compound compared to the other two borides. Fig. 3 (b) shows the charge density mapping (CDM) (in the units of e/Å$^3$) in the (101) crystallographic plane calculated to understand the bonding strength between different atoms. The atomic positions have been labeled in the figure. The replacement of M from Zr to Nb leads to a significant increase in the charge density at corresponding positions. The increased charge density results in stronger bonding between M atoms and B atoms owing to better hybridization between the states. The CDM also revealed that the bonding between M and B atoms are stronger than the bonding between S and B atoms. Based on the analyses of DOS and CDM, the largest values of stiffness constant, elastic moduli as well as hardness parameters are expected for the Nb$_2$SB compound, as reflected from the earlier calculated values. At the same time, we have performed the Mulliken population analysis to explain the chemical bonding associated within the compounds of interest.

*Mulliken atomic and bond population analysis*

The charge transfer mechanism can be understood using the atomic population analysis. For instance, in Table 4, the negative Mulliken charge is observed for S and B/C while it is positive for M (Zr, Hf and Nb), therefore, the charges are transferred from M to B/C and S atoms. This behavior is controlled by the electro negativities of the atomic species involved in the bonding. The presence of charge transfer disclosed the existence of ionic bonding in these compounds like many other MAX phase compounds. In fact, mixing of covalent and ionic bonding is an important characteristic feature of MAX phase materials. The Mulliken bond population analysis provides information regarding distribution of electron density in different bonds of the compounds. The bond-overlap population (BOP) analysis gives with quantitative measures of bonding and anti-bonding strengths [63, 66, 68]. The value of BOP *zero* indicates no significant interaction between electronic populations of the two atoms involved. The *positive* BOP value means the neighbor atoms are bonded whereas *negative* BOP value represents they are anti-bonded in the compounds. A high value of BOP indicates high degree of covalency of the bonds. The obtained values of BOP (Table 4) suggested that the bonds between M and B/C atoms exhibit higher degree of covalency as well as bonding strength than that between M and S atoms within these compounds which are in good agreement with bond hardness and CDM results found in preceding sections.

**Table 4 - Mulliken atomic and band overlap populations (BOP) of $M_2SB$ (M = Zr, Hf and Nb) compounds under study in comparison to $M_2SC$ (M = Zr, Hf and Nb).**

| Phases | Atoms | \multicolumn | | | | | | | | |
|---|---|---|---|---|---|---|---|---|---|---|

| | | \multicolumn{9}{c}{Mulliken atomic and band overlap populations} |

| Phases | Atoms | $s$ | $p$ | $d$ | Total | Charge (e) | Bond | Bond number $n^{\mu}$ | Bond overlap population $P^{\mu}$ |
|---|---|---|---|---|---|---|---|---|---|
| | B | 1.19 | 2.55 | 0.00 | 3.74 | -0.74 | B-Zr | 4 | 1.22 |
| $Zr_2SB$ | S | 1.79 | 4.56 | 0.00 | 6.35 | -0.35 | S-Zr | 4 | 0.76 |
| | Zr | 2.27 | 6.46 | 2.72 | 11.45 | 0.55 | | | |
| | C | 1.49 | 3.30 | 0.00 | 4.79 | -0.79 | C-Zr | 4 | 1.07 |
| $Zr_2SC$ | S | 1.78 | 4.57 | 0.00 | 6.35 | -0.35 | S-Zr | 4 | 0.73 |
| | Zr | 2.26 | 6.44 | 2.73 | 11.43 | 0.57 | | | |
| | B | 1.24 | 2.58 | 0.00 | 3.82 | -0.82 | B-Hf | 4 | 1.44 |
| $Hf_2SB$ | S | 1.82 | 4.60 | 0.00 | 6.42 | -0.42 | S-Hf | 4 | 0.81 |
| | Hf | 0.42 | 0.14 | 2.83 | 3.38 | 0.62 | | | |
| | C | 1.54 | 3.33 | 0.00 | 4.87 | -0.87 | C-Hf | 4 | 1.28 |
| $Hf_2SC$ | S | 1.81 | 4.60 | 0.00 | 6.41 | -0.41 | S-Hf | 4 | 0.80 |
| | Hf | 0.39 | 0.13 | 2.84 | 3.36 | 0.64 | | | |
| | B | 1.11 | 2.51 | 0.00 | 3.62 | -0.62 | B-Nb | 4 | 1.12 |
| $Nb_2SB$ | S | 1.77 | 4.40 | 0.00 | 6.17 | -0.17 | S-Nb | 4 | 0.79 |
| | Nb | 2.24 | 6.37 | 3.99 | 12.61 | 0.39 | | | |
| | C | 1.44 | 3.24 | 0.00 | 4.68 | -0.68 | C-Nb | 4 | 0.99 |
| $Nb_2SC$ | S | 1.77 | 4.44 | 0.00 | 6.21 | -0.21 | S-Nb | 4 | 0.73 |
| | Nb | 2.25 | 6.37 | 3.93 | 12.55 | 0.45 | | | |

### 3.3 The elastic anisotropy

Some of the physical processes such as unusual phonon modes, dislocation dynamics, precipitation, plastic deformation in solids, micro-scale crack creation etc. are influenced by mechanical anisotropy in solids [69]. Knowledge regarding anisotropy helps to enhance the mechanical stability of the material in any application [70]. These motivate us to study the mechanical anisotropy of 211 MAX phase borides and carbides. The $M_2SX$ (M = Zr, Hf and Nb; X = B and C) phases considered here are found to exhibit anisotropic features. In order to visualize the anisotropy of Young's modulus, compressibility, shear modulus and Poisson's ratio, we have shown in Figs. 4-6 their direction dependence by plotting (3D and 2D) their values using the ELATE code [71]. Only data for the titled boride MAX compounds are illustrated.

The spherical (in case of 3D) and circular (in case of 2D) nature of elastic moduli represents the isotropic nature of solids while deviation from spherical/circular symmetry indicates the anisotropy. The degree of anisotropy is measured by the amount of deviation from the perfect spherical/circular shape. Figs. 4(a), 5(a), and 6(a) show the anisotropy in Young's modulus ($Y$), from which it is observed that $Y$ is found to be isotropic in the $xy$ plane while it is anisotropic in $xz$ and $yz$ planes. $Y$ is minimum on the axes of both $xz$ and $yz$ planes and maximum at 45° angle of the axes on both $xz$ and $yz$ planes. Figs.4(b), 5(b), and 6(b) show the anisotropy in compressibility ($K$) in which similar anisotropic nature as for $Y$ is observed. The compressibility in the $xy$ plane exhibits isotropic character while in the $xz$ and $yz$ planes, it exhibits anisotropic nature with minimum values on the axes and maximum values at an angle of 45° of both the axes. It is observed in Figs.4(c), 5(c), and 6(c) that shear modulus shows opposite features as compared to $Y$ and $K$, exhibiting maximum on the axes in both $xz$ and $yz$ planes and minimum at an angle of 45° of both axes for Zr$_2$SB. The minimum values shift towards horizontal-axis for Hf$_2$SB and Nb$_2$SB. It is likely to be isotropic in $xy$ plane. Figs. 4(d), 5(d), and 6(d) exhibit the anisotropy of Poisson's ratio (v) in which very intricate anisotropic nature is observed in $xz$ and $yz$ planes. It is also likely to be isotropic in the $xy$ plane. In addition, the minimum and maximum values of $Y$, $K$, $G$ and $v$ of M$_2$SX (M = Zr, Hf and Nb; X = B and C) MAX phases are listed in Table 5 from which it is noted that the Nb$_2$SB are qualitatively more anisotropic compared to other two borides and Hf$_2$SB exhibits less anisotropy. It is obvious that the studied borides as well as carbides are elastically anisotropic in nature.

**Table 5 - The minimum and the maximum values of the Young's modulus, compressibility, shear modulus, and Poisson's ratio of M$_2$SX (M = Zr, Hf and Nb; X = B and C) MAX phases.**

| Phases | $Y_{min.}$ (GPa) | $Y_{max.}$ (GPa) | $A_Y$ | $K$ ($TPa^{-1}$) | $K$ ($TPa^{-1}$) | $A_K$ | $G_{min.}$ (GPa) | $G_{max.}$ (GPa) | $A_G$ | $v_{min.}$ | $v_{max.}$ | $A_v$ |
|---|---|---|---|---|---|---|---|---|---|---|---|---|
| Zr$_2$SB | 216.73 | 260.24 | 1.20 | 2.20 | 2.41 | 1.09 | 85.31 | 115.09 | 1.35 | 0.120 | 0.286 | 2.37 |
| Zr$_2$SC | 250.08 | 306.39 | 1.22 | 1.78 | 2.12 | 1.19 | 101.19 | 137.27 | 1.35 | 0.110 | 0.290 | 2.53 |
| Hf$_2$SB | 245.76 | 291.39 | 1.18 | 1.97 | 2.23 | 1.13 | 99.78 | 130.01 | 1.30 | 0.120 | 0.271 | 2.26 |
| Hf$_2$SC | 255.03 | 327.56 | 1.28 | 1.59 | 1.97 | 1.24 | 98.48 | 148.95 | 1.51 | 0.099 | 0.327 | 3.28 |
| Nb$_2$SB | 251.84 | 342.40 | 1.36 | 1.58 | 1.89 | 1.19 | 101.05 | 151.22 | 1.49 | 0.126 | 0.336 | 2.66 |
| Nb$_2$SC | 217.27 | 297.02 | 1.36 | 1.32 | 1.88 | 1.43 | 88.96 | 128.57 | 1.50 | 0.154 | 0.393 | 2.55 |

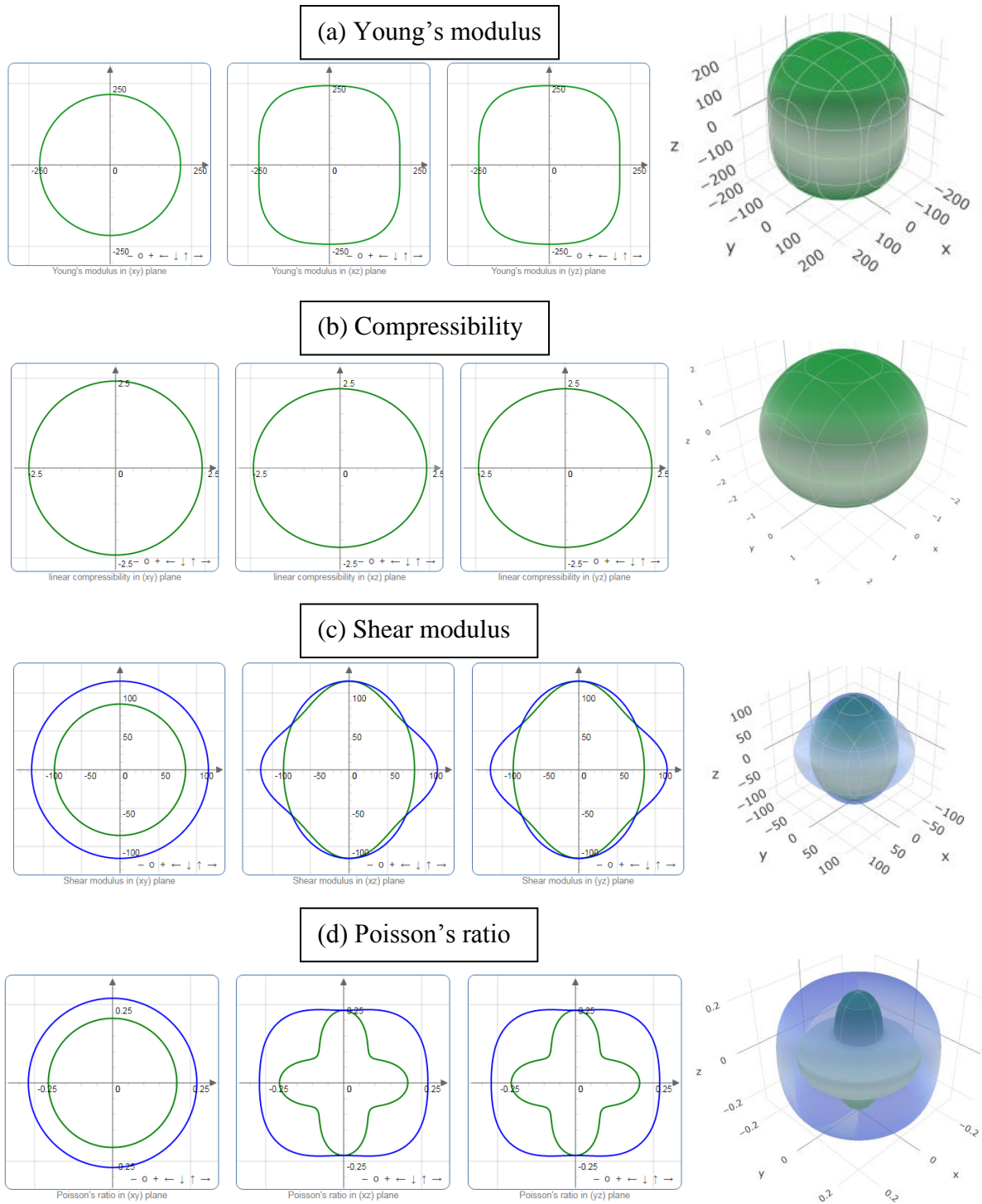

**Fig. 4 - The 2D and 3D plots of (a) Y, (b) K, (c) G and (d) υ of Zr₂SB.**

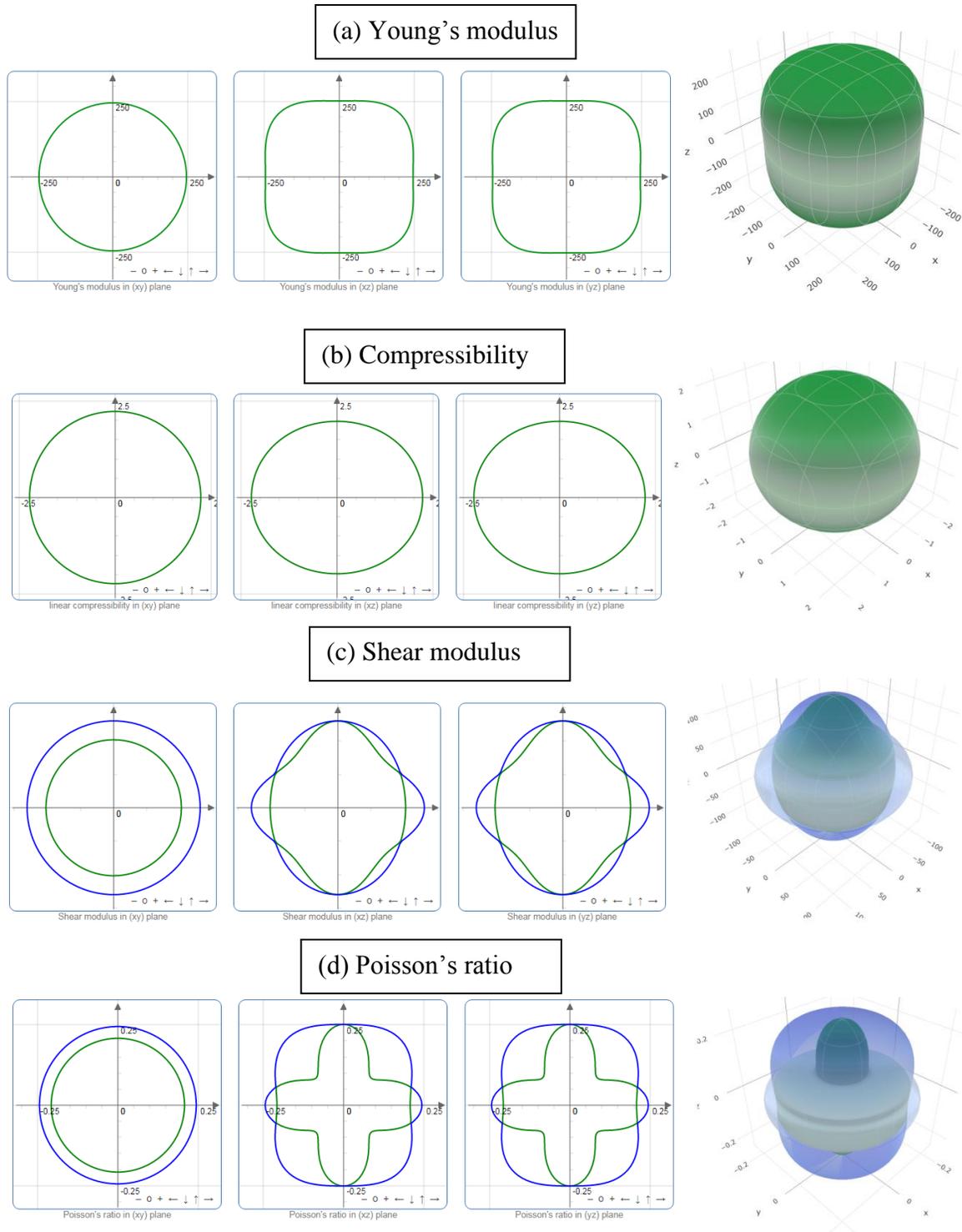

**Fig. 5 - The 2D and 3D plots of (a) Y, (b) K, (c) G and (d) υ of Hf₂SB.**

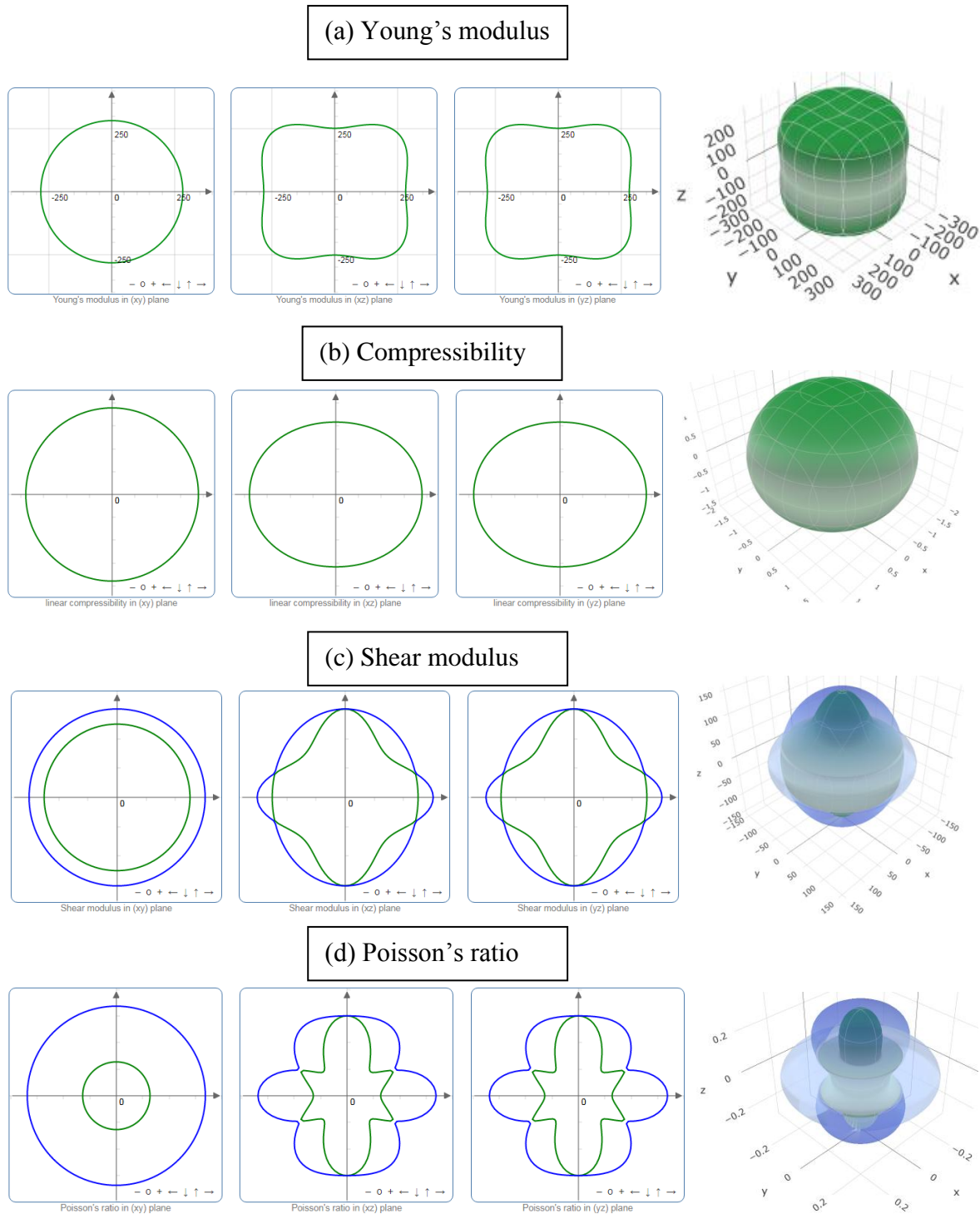

**Fig. 6 - The 2D and 3D plots of (a) Y, (b) K, (c) G and (d) υ of Nb₂SB.**

Moreover, we have also calculated some other anisotropic indices. For example, three shear anisotropic factors have been calculated using the following relations:

$$A_1 = \frac{1/6(C_{11}+C_{12}+2C_{33}-4C_{13})}{C_{44}}, A_2 = \frac{2C_{44}}{C_{11}-C_{12}}, A_3 = A_1 \cdot A_2 = \frac{1/3(C_{11}+C_{12}+2C_{33}-4C_{13})}{C_{11}-C_{12}}$$ [72] for the

{100}, {010} and {001} planes in between $\langle 011 \rangle$ and $\langle 010 \rangle$, $\langle 101 \rangle$ and $\langle 001 \rangle$, and $\langle 110 \rangle$ and $\langle 010 \rangle$ directions, respectively. The listed values of $A_i$'s presented in Table 6 which revealed that the studied MAX phases are anisotropic ($A_i = 1$ implies the isotropic nature) in nature. The elastic anisotropy in the bulk modulus $B_a$ and $B_c$ along the $a$- and $c$-axes are calculated using the following relations [73]: $B_a = a\frac{dP}{da} = \frac{\Lambda}{2+\alpha}, B_c = c\frac{dP}{dc} = \frac{B_a}{\alpha}$, where $\Lambda = 2(C_{11} + C_{12}) + 4C_{13}\alpha + C_{33}\alpha^2$ and $\alpha = \frac{(C_{11}+C_{12})-2C_{13}}{C_{33}+C_{13}}$. The bulk modulus along $a$-axis and $c$-axis is different (Table 6), indicating the anisotropic nature of the bulk modulus. In addition, the ratio of linear compressibility coefficients ($k_c/k_a$) along the $a$- and $c$-axis, respectively is calculated using the relation [74]: $\frac{k_c}{k_a} = C_{11} + C_{12} - 2C_{13}/(C_{33} - C_{13})$. The listed values of $k_c/k_a$ ratio in Table 6 are lower than unity (1), revealing lower compressibility along $c$- than along $a$-axis for all the phases under study.

**Table 6 - Anisotropic factors, $A_1$, $A_2$, $A_3$, $k_c/k_a$, $B_a$, $B_c$, percentage anisotropy factors $A_G$ and $A_B$, and universal anisotropic index $A^U$ of $M_2SX$ (M = Zr, Hf and Nb; X = B and C) MAX phases.**

| Phase | $A_1$ | $A_2$ | $A_3$ | $B_a$ | $B_c$ | $k_c/k_a$ | $A_B$ | $A_G$ | $A^U$ |
|-------|-------|-------|-------|-------|-------|-----------|-------|-------|-------|
| Zr$_2$SB | 0.85 | 1.35 | 1.14 | 412.9 | 453.5 | 0.91 | 0.000422 | 0.00906 | 0.092 |
| Zr$_2$SC | 0.73 | 1.34 | 0.98 | 470.2 | 556.4 | 0.85 | 0.001294 | 0.01055 | 0.109 |
| Hf$_2$SB | 0.76 | 1.27 | 0.97 | 447.2 | 506.2 | 0.88 | 0.000674 | 0.00788 | 0.080 |
| Hf$_2$SC | 0.65 | 1.39 | 0.90 | 505.5 | 627.3 | 0.81 | 0.001688 | 0.01714 | 0.178 |
| Nb$_2$SB | 0.62 | 1.26 | 0.78 | 511.8 | 646.1 | 0.79 | 0.001128 | 0.01475 | 0.152 |
| Nb$_2$SC | 0.63 | 1.19 | 0.75 | 529.9 | 755.7 | 0.70 | 0.003244 | 0.01527 | 0.161 |

The percentage anisotropies in compressibility and shear modulus are calculated using the relations [75]: $A_B = \frac{B_V - B_R}{B_V + B_R} \times 100\%$ and $A_G = \frac{G_V - G_R}{G_V + G_R} \times 100\%$, where the subscripts denote the upper bound (Voigt, $V$) and lower bound (Reuss, $R$) of $B$ and $G$. The upper bound (Voigt, $V$) and lower bound (Reuss, $R$) of $B$ and $G$ are also used to calculate the universal anisotropic index $A^U$, where $A^U = 5\frac{G_V}{G_R} + \frac{B_V}{B_R} - 6 \geq 0$. The non-zero value of $A^U$ implies the anisotropic nature of solids [76]. The obtained non-zero values of $A^U$ revealed the anisotropic nature of the studied borides. At the end of this section we can conclude that $M_2SX$ (M = Zr, Hf and Nb; X = B and C) phases are anisotropic in nature as confirmed from the Table 6.

### 3.4 Optical properties

It already has been mentioned that the MAX phases materials are potential candidates to be used as in optical appliances. Therefore, the recently synthesized borides are also expected to be suitable for possible applications. With an intention to disclose the optical response of these borides, we have calculated several optical constants for the first time. The optical behavior in the IR, visible and UV region is significant from optoelectronic applications point of view. The details of formalisms and methods employed to study the optical properties can be found elsewhere [77, 78].

The dielectric functions of metallic compounds are attributed to the electronic transitions (both inter-band and intra-band). Introduction of Drude term is mandatory for the investigation of the dielectric function of metallic solids, usually done by introducing the plasma frequency and a broadening factor during first principles calculations [65, 66]. In our present case, the borides are metallic and we have used a damping of 0.05 eV and unscreened plasma frequency of 3 eV that leads the enhanced lower energy region of the obtained spectra. With an expectation to get more effective k-points on the Fermi surface, a Gaussian smearing (0.5 eV) was used. The calculated optical constants of$M_2SB$ (M = Zr, Hf and Nb) borides have been displayed for photon energies up to 25 eV. The optical constants of $M_2SC$ (M = Zr, Hf and Nb) carbides are also calculated to compare with those of the corresponding borides.

Fig. 7 (a) shows the real part $\varepsilon_1(\omega)$ of the dielectric function in which the peaks in the low energy are attributed from the intra-band transitions of electrons [79]. The large negative value of dielectric constant $\varepsilon_1(\omega)$ indicates the Drude-like behavior which is common for metallic system. The $\varepsilon_1(\omega)$ is found to cross zero value from below (negative values) at 14. 49 eV, 14.96 eV and 15.37 eV for $M_2SB$ (M = Zr, Hf and Nb) borides, respectively and for 15.36 eV, 16.44 eV and 16.44 eV for $M_2SC$ (M = Zr, Hf and Nb) carbides, respectively. These are the characteristic points at which a sharp drop in the reflectivity spectrum (Fig. 7 (g)) and the peaks in energy loss function (Fig. 7 (h) are observed. Fig. 7 (b) shows the imaginary part, $\varepsilon_2(\omega)$, of the dielectric function $\varepsilon(\omega)$ of $M_2SB$ (M = Zr, Hf and Nb) borides along with the $\varepsilon_2(\omega)$ of $M_2SC$ (M = Zr, Hf and Nb) carbides. The $\varepsilon_2(\omega)$ measures the photon absorption in which the peaks are associated with electron excitation (inter-band transitions) owing to photon absorption. The intra-band transition of electrons takes place within M-$d$ states and inter-band transition of electrons takes place from valence band to conduction band from S-$p$ states, respectively. The metallic nature of $M_2SB$ (M = Zr, Hf and Nb) borides is also revealed from the coincidence of $\varepsilon_1(\omega)$ and $\varepsilon_2(\omega)$ at the energy where $\varepsilon_2(\omega)$ approached zero from above and the $\varepsilon_1(\omega)$ crossed zero from below.

Fig. 7 (c) shows the refractive index, $n(\omega)$,of $M_2SB$ (M = Zr, Hf and Nb) borides. This is an important optical constant useful to design optical devices, for instance photonic crystals, wave guides etc. The static values of $n(0)$ of $M_2SB$ (M = Zr, Hf and Nb) borides are 29, 12 and 8, respectively. On the other hand, the static values of $n(0)$ of $M_2SC$ (M = Zr, Hf and Nb) carbides are 20, 18 and 12, respectively. The extinction coefficients, $k(\omega)$, of $M_2SX$ (M = Zr, Hf and Nb, X = B and C) MAX phases are displayed in Figs. 7 (d). The loss of electromagnetic radiation by absorption is assessed by $k(\omega)$ which is found to vary in a similar manner of $\varepsilon_2(\omega)$.

The absorption coefficient, $\alpha(\omega)$, quantifies the ability of a material to absorb incident radiation in different spectral range and provides with the information regarding its ability to convert solar energy in other useful forms. The absorption coefficient of $M_2SX$ (M = Zr, Hf and Nb, X = B and C) MAX phases is demonstrated in Fig. 7 (e) in which the spectra are found to rise from zero photon energy because of the metallic nature of studied compounds. The first peaks are observed in the range of 1.0 eV to 2.5 eV and then reach to a maximum in the energy range of 7.0 eV to 9.0 eV. The spectra are almost similar for carbides and borides except slight differences in the positions and heights of the peaks. However, the absorption coefficients are quite high in the high energy range, revealing that $M_2SX$ (M = Zr, Hf and Nb, X = B and C) MAX phases are promising absorbing materials within this energy range. This type of materials is widely used in optical and optoelectronic devices in the visible and ultraviolet energy regions. The photoconductivity also starts from zero photon energy as shown in Fig. 7(f). The spectral behavior of photoconductivity confirms the good metallic nature of these compounds. The absorption coefficient and photoconductivity features are in good agreement with the electronic DOS results.

The reflectivity curves of $M_2SX$ (M = Zr, Hf and Nb, X = B and C) MAX phases are also shown in Fig. 7 (g). The curves start with 0.86, 0.71 and 0.59 for $Zr_2SB$, $Hf_2SB$ and $Nb_2SB$ while the starting values for corresponding carbides are 0.82, 0.79 and 0.72 for $Zr_2SC$, $Hf_2SC$ and $Nb_2SC$, respectively. According to Li et al. [80, 81] a MAX compound will be capable of reducing solar heating if it has reflectivity ~ 44% in the visible region. Among the considered borides and carbides, $Zr_2SB$ has highest starting value, highest minimum value (43%) and the lowest energy range (2.74 − 3.8 eV) in which the reflectivity value is lower than 44%, therefore, it will be the best candidate for shielding purpose to reduce solar heating. $Nb_2SB$ also shows promising spectral features for the mentioned purpose. $Hf_2SB$ will be the worst one for shielding purpose for the same. The studied carbides can also be ranked as $Zr_2SC$ ← $Nb_2SC$ ← $Hf_2SC$ in consideration of their reflectivity characteristics.

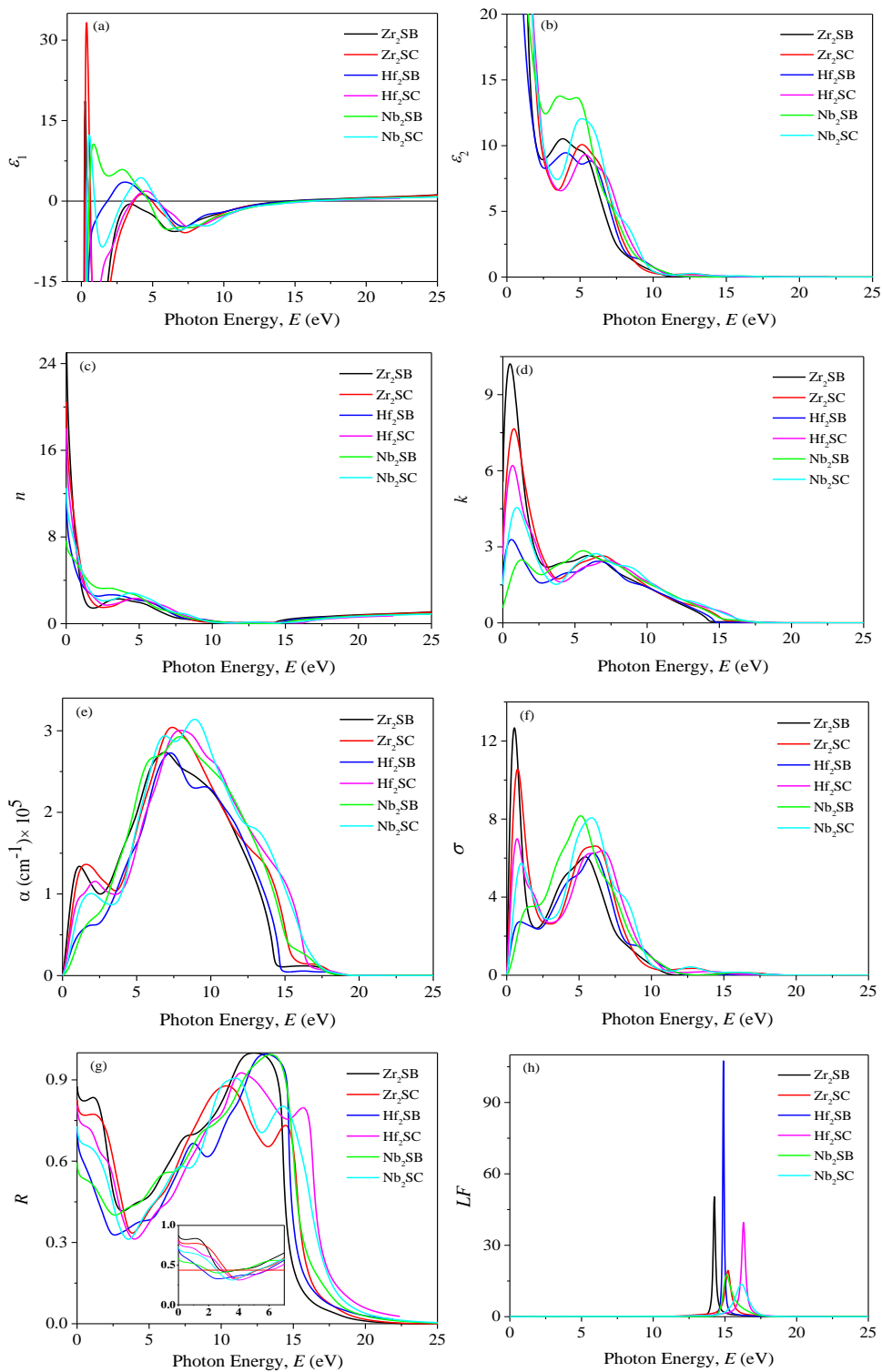

**Fig. 7 -** (a) real part $\varepsilon_1$ and (b) imaginary part of dielectric function $\varepsilon_2$, (c) refractive index *n*, (d) extinction coefficient *k*, (e) photoconductivity $\sigma$, (f) absorption $\alpha$, (g) reflectivity *R*, and (h) loss function *LF* of $M_2SX$ (M = Zr, Hf and Nb; X = B and C) MAX phases as function of photon energy.

The energy loss of electrons traversing the materials is common and this loss is measured by an optical constant known as the loss function. The calculated loss functions of the titled compounds are displayed in Fig. 7 (h). The peak frequency in the loss function, known as the plasma frequency ($\omega_p$), is observed at 14.26, 14.9 and 15.05 eV for $Zr_2SB$, $Hf_2SB$ and $Nb_2SB$, respectively, while the peaks are at 15.21, 16.18 and 16.33 eV for $Zr_2SC$, $Nb_2SC$ and $Hf_2SC$, respectively. The characteristic frequency in loss function defines an energy at which $\varepsilon_1(\omega)$ goes through zero from below while $\varepsilon_2(\omega)$ goes through zero from above. Moreover, the peak energy of the loss function is associated with the trailing edges of the reflection spectrum. Above this peak energy the compounds under study become transparent to incident radiation.

### *3.5 The dynamical stability*

The dynamical stability of a material is an important criterion that determines the applicability of the material in practical situations where external stimuli are time dependent. The titled borides are synthesized at high temperature [37, 39]. The materials synthesized at any temperature may not be stable at all conditions. For example, the materials synthesized at high temperature may also not be stable at low temperature [82]. The lattice vibration or phonon frequency depends on the temperature of the crystal lattice. In order to check the stability of the synthesized borides at low temperature, we have calculated the phonon dispersion curve (PDC) at absolute zero using the Density Functional Perturbation Theory (DFPT) linear-response method [83, 84]. The calculated phonon dispersion curves and the total phonon density of states (PHDOS) of $M_2SB$ (M = Zr, Hf and Nb) borides along the high symmetry direction of the crystal Brillouin zone (BZ) are displayed in Figs. 8 (a-c). The stability is checked by the phonon frequency over the whole BZ; the positive frequencies suggest the stability while the negative frequencies at any *k*-points indicate the instability of the compounds. As there is no negative frequency in the displayed PDCs in Figs. 8 (a-c), therefore, the titled borides are dynamically stable. In addition, some other information can also be found from the PDCs. The eight atoms in the unit cell of studied borides lead to 24 vibrational modes in the PDCs, three of which are acoustic modes and the rest (21) are optical modes. The lowest three modes belong to acoustic branch in which the dispersion curve is of the form $\omega = vk$ at low values of *k*, representing the $\omega(k)$ relationship of the sound waves. The optical branch is formed by the upper vibrational modes. The optical phonons are generated owing to the out-of-phase oscillations of the atoms due to photon induced excitation. The frequency of acoustic modes at the G point is zero. A overlapping between acoustic modes and lower optical branches results in no phononic band gap between the two branches.

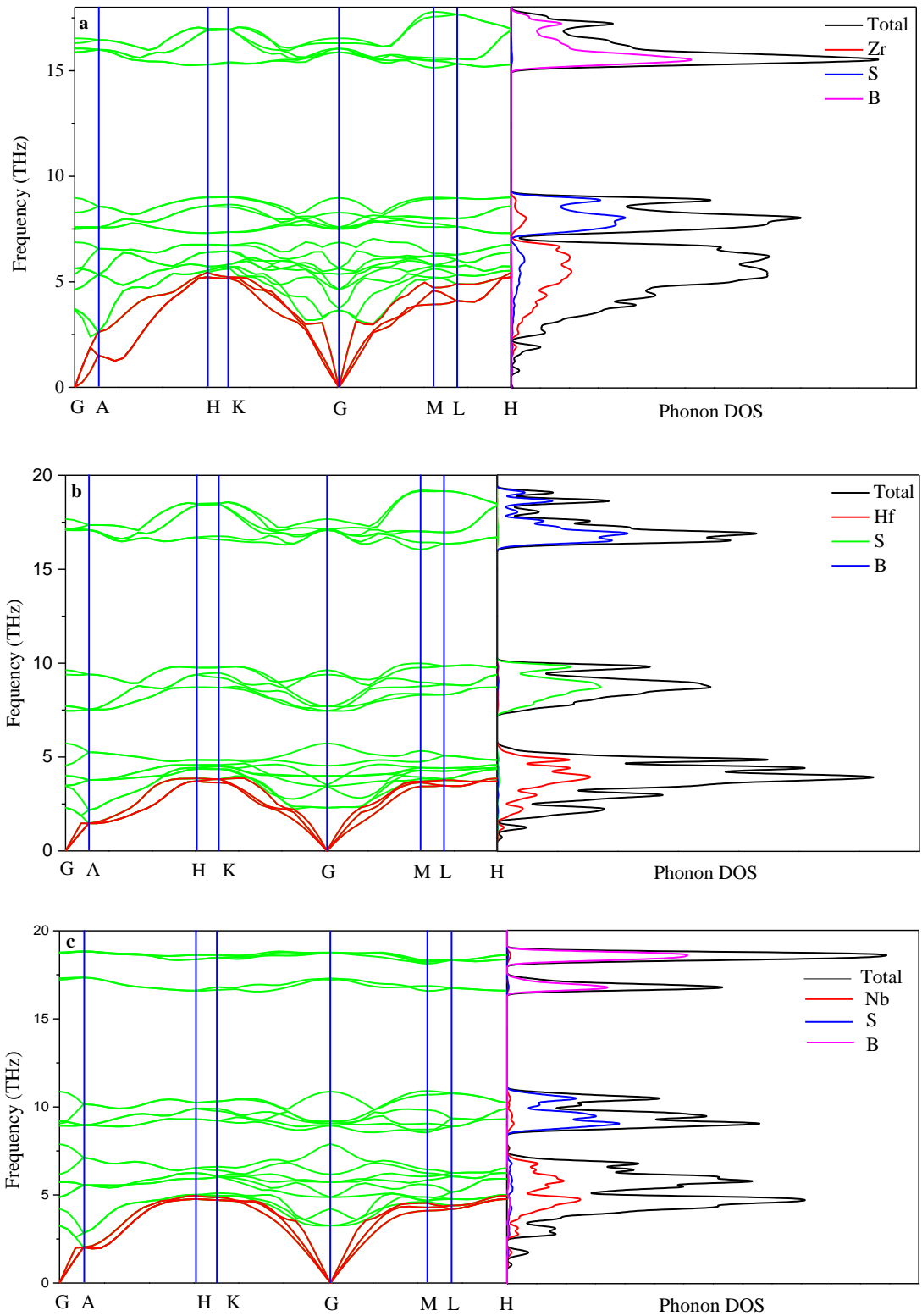

**Fig. 8 - Phonon dispersion curve and phonon DOS of (a) Zr₂SB, (b) Hf₂SB and (c) Nb₂SB.**

The phonon DOS is also shown in association with the PDC in which the sharp peaks correspond to the flat modes of the phonon dispersion curves and upturn or downturn of the dispersion leads reduction of the peaks. The contribution from different atoms is also shown in the partial phonon DOS in which it is clear that the top most optical modes are credited to the B atoms. The middle dispersions of the optical branches are credited to the S atoms. The modes in the lower optical branches are credited to the M atoms. The acoustic modes are attributed to the M atoms.

### 3.6 Thermal properties

*Lattice thermal conductivity*

Heat energy carried due to temperature gradient by lattice vibration in a solid can be measured by the lattice thermal conductivity ($k_{ph}$). Slack [85] derived an empirical formula to estimate the $k_{ph}$ as follows:

$$k_{ph} = A(\gamma)\frac{M_{av}\Theta_D^3\delta}{\gamma^2 n^{2/3}T}$$

$$\gamma = \frac{3(1+v)}{2(2-3v)}$$

Here $M_{av}$ be the average atomic mass per atom in a compound, $\Theta_D$ is the Debye temperature, $n$ is the total number of atoms in the unit cell, $T$ is the absolute temperature and $\gamma$ is the Grüneisen parameter. $A(\gamma)$ is a function of $\gamma$. Once the Grüneisen parameter is calculated, we can estimate the coefficient $A(\gamma)$ as follows:

$$A(\gamma) = \frac{4.85628 \times 10^7}{2(1-\frac{0.514}{\gamma}+\frac{0.228}{\gamma^2})}$$

The minimum thermal conductivity, $k_{min}$, of a material can be estimated using modified Clarke's model [86] as follows: $k_{min} = k_B v_m \left(\frac{M}{n\rho N_A}\right)^{-\frac{2}{3}}$

The theoretical description and analytical equations to calculate the $\Theta_D$, longitudinal wave velocity ($v_l$), transverse wave velocity ($v_t$), average sound velocity ($v_m$), and melting temperature ($T_m$) of a hexagonal system can be found in previous reports [23,64]. Herein, all the technologically important parameters such as $k_{ph}, k_{min}, \gamma, \Theta_D, v_l, v_t,$ and $T_m$ were calculated using the aforementioned formulae and are tabulated in Table 7. The temperature dependent $k_{ph}$ for M$_2$SX (M = Zr, Hf and Nb; X = B and C) MAX phases are shown in Fig. 9.

According to the Clarke's equation, $k_{min}$ is strongly dependent on the average sound velocity ($v_m$). The variation of $k_{min}$ for all the studied compounds is related to the $v_m$. Among all borides and carbides, Hf$_2$SB compound shows the lowest value of $k_{min}$ due to the low value of $v_m$. We have studied the lattice thermal conductivity up to a temperature of 1800 K considering the melting temperatures of the compounds as shown in Table 7. The phonon thermal conductivity for M$_2$SX (M = Zr, Hf and Nb; X = B and C) compounds exhibits a decreasing trend with temperature which can be expressed as: 10749.44/T, 12556.49/T, 9592.569/T, 6977.325/T, 12926.19/T and 9390.386/T, respectively. At higher temperature $k_{ph}$ shows a tendency of saturation. The values of total thermal conductivity of all MAX phase compounds at room temperature are usually in the range of from 12 to 60 W/mK [7]. It is noticed from Table 7 that comparatively high value of $k_{ph}$, ranging from 23.25 − 43.08 W/mK, for all 211 MAX phase borides and carbides under study at room temperature (300 K) are found, although the electronic contribution to the thermal conductivity has not been included. Altogether this suggests that the titled compounds possess relatively high thermal conductivities among the existing MAX phase nanolaminates. However, the highest value (43.08 W/mK) of $k_{ph}$ was obtained for Nb based boride material as the bonding strength and $\Theta_D$ of this particular compound are the highest among all the borides under study as described in Section 3.2 and shown in Fig. 3 (a).

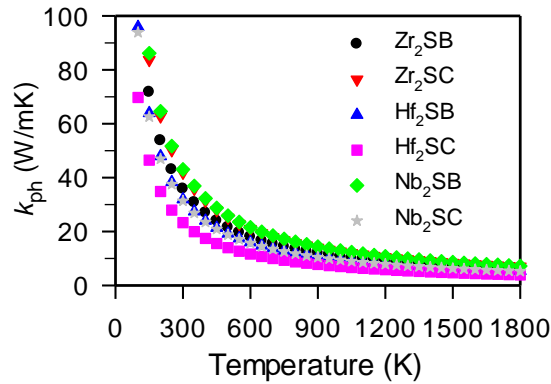

**Fig. 9 - Temperature dependence of calculated lattice thermal conductivity, $k_{ph}$ (W/mK) for M$_2$SX (M = Zr, Hf and Nb; X = B and C) MAX phases.**

**Table 7 - Calculated crystal density, longitudinal, transverse and average sound velocities ($v_l$, $v_t$, and $v_m$), Debye temperature, $\Theta_D$, minimum thermal conductivity, $K_{min}$, lattice thermal conductivity, $k_{ph}$ at 300 K and Grüneisen parameter, $\gamma$ for M$_2$SX (M = Zr, Hf and Nb; X = B and C) MAX phases.**

| Phases | $\rho$ (kg/m$^3$) | $v_l$(m/s) | $v_t$(m/s) | $v_m$(m/s) | $\Theta_D$ (K) | $K_{min}$ (W/mK) | $k_{ph}^*$(W /mK) | $\gamma$ | $T_m$(K) |
|---|---|---|---|---|---|---|---|---|---|
| Zr$_2$SB | 5676 | 6947 | 4172 | 4615 | 540 | 0.98 | 35.83 | 1.32 | 1542 |
| Zr$_2$SC | 5514 | 7610 | 4567 | 5052 | 604 | 1.12 | 41.85 | 1.36 | 1712 |
| Hf$_2$SB | 10198 | 5465 | 3309 | 3657 | 430 | 0.79 | 31.97 | 1.32 | 1656 |
| Hf$_2$SC | 10192 | 5784 | 3431 | 3800 | 454 | 0.85 | 23.25 | 1.60 | 1778 |
| Nb$_2$SB | 6701 | 7294 | 4366 | 4831 | 594 | 1.14 | 43.08 | 1.36 | 1824 |
| Nb$_2$SC | 6310 | 7308 | 4079 | 4542 | 565 | 1.09 | 31.30 | 1.40 | 1790 |

*Heat capacities and thermal expansion coefficient*

The phonon heat capacity at constant volume ($C_v$) in the temperature ranging from 0 to 1000 K, can be estimated using the quasi-harmonic Debye model as follows[87-90]:

$$C_v = 9nN_A k_B \left(\frac{T}{\Theta_D}\right) \int_0^{x_D} dx \, \frac{x^4}{(e^x - 1)^2}$$

where, $x_D = \frac{\Theta_D}{T}$ ; and $n$ be the number of atoms per formula unit, $N_A$ be the Avogadro's number and $k_B$ be the Boltzmann constant. Again, the linear thermal expansion coefficient (*TEC*), specific heat at constant pressure ($C_p$) are calculated by the following equations [87, 91]:

$$TEC = \frac{\gamma C_v}{3B_T v_m}$$

$$C_p = C_v[1 + (TEC)\gamma T]$$

where, $B_T, v_m$ and $\gamma$ are the isothermal bulk modulus, molar volume and Grüneisen parameter, respectively.

The specific heats, $C_v$, $C_p$ and TEC for M$_2$SX (M = Zr, Hf and Nb; X = B and C) compounds as a function of temperature are shown in Figs. 10 (a-c) . The specific heats $C_v$ and $C_p$ are found to increase rapidly with the increase in temperature up to 300 K, which support the Debye T$^3$ power-law. Both $C_v$ and $C_p$ increase slowly at higher temperature and finally take up the Dulong-Petit limit value ($3nN_A k_B$).

The thermal expansion of materials is caused by anharmonicity in the lattice dynamics and is responsible for the difference between the specific heats $C_p$ and $C_v$. The variations of *TEC* with temperature is shown in Fig. 10 (c) for M$_2$SX (M = Zr, Hf and Nb; X = B and C) compounds. *TEC*(T) followed similar trend of heat capacities ($C_v$ and $C_p$). The values of *TEC* for M$_2$SX (M = Zr, Hf and Nb; X = B and C) compounds are 7.0 × 10$^{-6}$, 6.71 × 10$^{-6}$, 5.90 × 10$^{-6}$, 6.22 × 10$^{-6}$, 7.05 × 10⁻⁶ and 6.30 × 10⁻⁶ (all in unit of K⁻¹) at room temperature (300 K), respectively. However, the calculated $K_{min}$, *TEC* and $T_m$ values for Hf₂SB and Hf₂SC are comparable with a promising typical thermal barrier coating (TBC) material Y₄Al₂O₉ [92, 93] and also comparable with some other related potential TBC compounds [16, 94]. Therefore, it can be concluded that the low minimum thermal conductivity, $K_{min}$, low linear thermal expansion coefficient, *TEC*, moderately damage tolerant behaviour and high melting point, $T_m$ indicate that the Hf based boride and carbide 211 MAX compounds can be possible candidates as thermal barrier coating materials for high temperature applications.

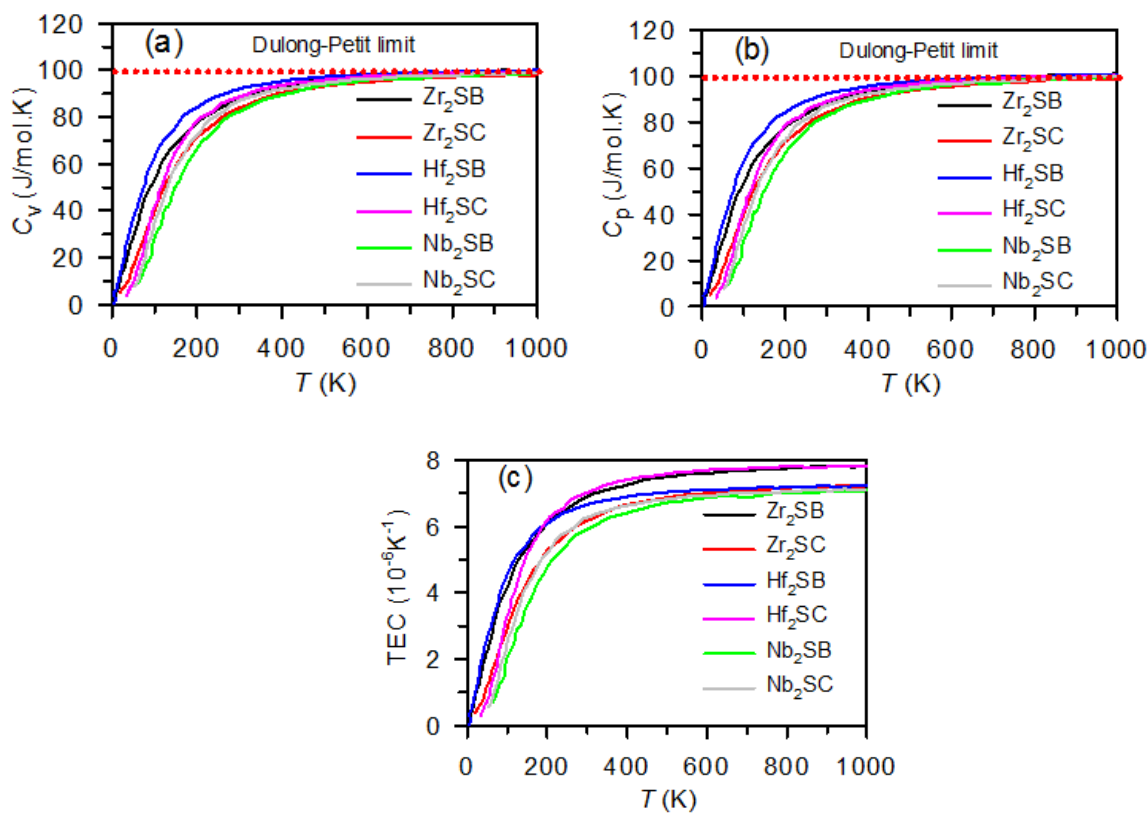

**Fig. 10 - Vibrational heat capacity (a) $C_v$ at constant volume, (b) $C_p$ at constant pressure and (c) linear thermal expansion coefficient (*TEC*) for M₂SX (M = Zr, Hf and Nb; X = B and C) MAX phases.**

## 4. Conclusions

A systematic study of 211 MAX phase borides and carbides of M₂SX (M = Zr, Hf and Nb; X = B and C) compounds was conducted based on the density functional theory. A number of hitherto unexplored properties such as micro- and macro-hardness, Vickers hardness, elastic anisotropy indices, machinability index and Cauchy pressure were investigated. Optical, thermal and lattice dynamical properties are thoroughly studied and explained in detail also for the first time. Electronic density of states was revisited. All results are compared with available data and are found in very good agreement. The values of $B$, $G$, $Y$, $H_{\text{micro}}$, $H_{\text{macro}}$ and $H_{\text{v}}$ in the $M_2SB$ ($M$ = Zr, Hf and Nb) borides can be ranked as follows: $Nb_2SB > Hf_2SB > Zr_2SB$. Meanwhile, the compound $Nb_2SB$ exhibits the best mechanical properties among the all studied borides and carbides. The origin of the variation of mechanical properties has been discussed based on density of states and mapping of charge density distribution. The Mulliken population analysis describes the mechanism of charge transfer and bonding types (covalent and ionic). The optical absorption as well as the photoconductivity spectra reveals the metallic nature of all boride compounds. No imaginary frequency has been observed in phonon dispersion curves implying the dynamical stability of all the titled boride MAX phases. Various thermal properties, like minimum thermal conductivity ($k_{min}$), Grüneisen parameter ($\gamma$), lattice thermal conductivity ($k_{ph}$), Debye temperature ($\Theta_D$), heat capacities ($C_p$, $C_v$), linear thermal expansion coefficient ($TEC$), and melting point ($T_m$) have been calculated. The order of $T_m$ of 211 MAX phase borides under study has the following sequence: $Hf_2SB < Zr_2SB < Nb_2SB$. This can be understood from the bonding strength order: $Hf_2SB < Zr_2SB < Nb_2SB$. In general, lower value of $\Theta_D$ and bonding strength indicates lower value of $k_{min}$. A decreasing trend of $k_{ph}$ for $M_2SX$ ($M$ = Zr, Hf and Nb; X = B and C) phases is obtained with temperature following 10749.44/T, 12556.49/T, 9592.569/T, 6977.325/T, 12926.19/T and 9390.386/T functional dependences, respectively. Low $K_{min}$ and $TEC$, moderately damage tolerant behavior and high melting point, $T_m$ indicate that the Hf based borides and carbides under study might be a potential candidate as thermal barrier coating material in high temperature applications.

**References**

[1]     Jeitschko W, Nowotny H, Benesovsky F. Carbides of formula T$_2$MC. J Less Common Metals 1964;7:133-138.

[2]     Wolfsgruber H, Nowotny H, Benesovsky F. Die Kristallstruktur von Ti$_3$GeC$_2$. Monatsh Chem 1967;98:2403-2405.

[3]     Nowotny H. Strukturchemie einiger Verbindungen der Übergangsmetalle mit den elementen C, Si, Ge, Sn. Prog. Solid State Chem 1970;2:27-70.

[4]     Nowotny H, Schuster JC, Rogl P., Structural chemistry of complex carbides and related compounds. J Solid State Chem 1982;44:126-133.

[5]     Barsoum MW, El-Raghy T. Synthesis and Characterization of a Remarkable Ceramic: Ti$_3$SiC$_2$. J Am Ceram Soc 1996;79:1953-1956.

[6]     Barsoum MW. The $M_{N+1}AX_N$ phases: A new class of solids: Thermodynamically stable nanolaminates. Prog Solid State Chem 2000;28:201-281.

[7]     Barsoum MW. MAX Phases: Properties of Machinable Ternary Carbides and Nitrides; WILEY-VCH Verlag GmbH & Co. KGaA, 2013.

[8]     Barsoum MW, El-Raghy T. The MAX Phases: Unique New Carbide and Nitride Materials. J Am Ceram Soc 2001;89:334-343.

[9]     Ali MA, Hossain MM, Jahan N, Islam AKMA, Naqib SH. Newly synthesized $Zr_2(Al_{0.58}Bi_{0.42})C$, $Zr_2(Al_{0.2}Sn_{0.8})C$, and $Zr_2(Al_{0.3}Sb_{0.7})C$ MAX phases: A first-principles study. Comput Mater Sci 2017;131:139-145.

[10]    Chakraborty P, Chakrabarty A, Dutta A, Dasgupta TS. Soft MAX phases with boron substitution: A computational prediction. Phys Rev Mater 2018;2:103605.

[11]    Horlait D, Middleburgh SC, Chroneos A, Lee WE. Synthesis and DFT investigation of new bismuth-containing MAX phases, Sci. Rep. 2016;6:18829.

[12]    Lapauw T, Lambrinou K, Cabioc'h T, Halim J, Lu J, Pesach A, et al. Synthesis of the new MAX phase $Zr_2AlC$. J Eur Ceram Soc 2016;36:1847–1853.

[13]    Lapauw T, Halim J, Lu J, Cabioc'h T, Hultman L, Barsoum MW, et al. Synthesis of the novel $Zr_3AlC_2$ MAX phase, J Eur Ceram Soc 2016;36:943-947.

[14]    Anasori B, Halim J, Lu J, Voigt CA, Hultman L, Barsoum MW. $Mo_2TiAlC_2$: A new ordered layered ternary carbide, Scr Mater 2015;101:5–7.

[15]    Hu C, Lai CC, Tao Q, Lu J, Halim J, Sun, et al. $Mo_2Ga_2C$: a new ternary nanolaminated carbide. Chem Commun 2015;51:6553.

[16]    Hadi MA, Kelaidis N, Naqib SH, Chroneos A, Islam AKMA. Mechanical behaviors, lattice thermal conductivity and vibrational properties of a new MAX phase $Lu_2SnC$. J Phys Chem Solids 2019;129:162-171.

[17]    Meshkian R, Tao Q, Dahlqvist M, Lu J, Hultman L, Rosen J. Theoretical stability and materials synthesis of a chemically ordered MAX phase, $Mo_2ScAlC_2$, and its two-dimensional derivate $Mo_2ScC_2$ MXene. Acta Mater 2017;125: 476-480.

[18]    Zapata-Solvas E, Hadi MA, Horlait D, Parfitt DC, Thibaud A, Chroneos A, et al. Synthesis and physical properties of $(Zr_{1-x},Ti_x)_3AlC_2$ MAX phases. J Am Ceram Soc 2017;100:3393-3401.

[19]    Horlait D, Grasso S, Nasiri N A, Burr PA, Lee WE. Synthesis and Oxidation Testing of MAX Phase Composites in the Cr–Ti–Al–C Quaternary System. J Am Ceram Soc 2016;99:682-690.

[20]    Ali MS, Rayhan MA, Ali MA, Parvin R, Islam AKMA. New MAX Phase Compound $Mo_2TiAlC_2$: First-principles Study. J Sci Res 2016;8:109-117.

[21]    Roknuzzaman M, Hadi MA, Ali MA, Hossain MM, Uddin MM, Alarco JA, et al. First hafnium-based MAX phase in the 312 family, Hf3AlC2: A first-principles study. J Alloys Compd 2017;727:616-626.

[22]    Ali MA, Khatun MR, Jahan N, Hossain MM. Comparative study of $Mo_2Ga_2C$ with superconducting MAX phase $Mo_2GaC$: A first-principles calculations. Chinese Phys B 2017;26:033102.

[23]    Ali MA, Hossain MM, Hossain MA, Nasir MT, Uddin MM, Hasan MZ, et al. Recently synthesized $(Zr_{1-x}Ti_x)_2AlC$ $(0 \leq x \leq 1)$ solid solutions: Theoretical study of the effects of M mixing on physical properties. J Alloys Compd. 2018;743:146-154.

[24]    Zhou AG, M.W. Barsoum, Kinking nonlinear elastic deformation of $Ti_3AlC_2$, $Ti_2AlC$, $Ti_3Al(C_{0.5},N_{0.5})_2$ and $Ti_2Al(C_{0.5},N_{0.5})$. J Alloys Compd 2010;498:62–70.

[25]    Radovic M, Ganguly A, Barsoum MW. Elastic properties and phonon conductivities of $Ti_3Al(C_{0.5},N_{0.5})_2$ and $Ti_2Al(C_{0.5},N_{0.5})$ solid solutions. J Mater Res 2008;23:1517–1521.

[26]    Manoun B, Saxena SK, Hug G, Ganguly A, Hoffman EN, Barsoum MW. Synthesis and compressibility of $Ti_3(Al_{0.8},Sn_{0.2})C_2$ and $Ti_3Al(C_{0.5},N_{0.5})_2$. J Appl Phys 2007;101:113523/1-113523/7.

[27]    Zhang Y, Yang S, Wang C. First-principles calculations for point defects in MAX phases $Ti_2AlN$. Mod Phys Lett B 2016;30:1650101.

[28]    Christopoulos SRG, Filippatos PP, Hadi MA, Kelaidis N, Fitzpatrick M E, Chroneos A. Intrinsic defect processes and elastic properties of $Ti_3AC_2$ (A = Al, Si, Ga, Ge, In, Sn) MAX phases. J Appl Phys 2018;123:025103.

[29]    Dahlqvist M. Benefits of oxygen incorporation in atomic laminates. J Phys: Condens Matter 2016;28:135501.

[30]    Ding H, Glandut N, Fan X, Liu Q, Shi Y, Jie J. First-principles study of hydrogen incorporation into the MAX phase $Ti_3AlC_2$. Int J Hydrogen Energy 2016;41:6387–6393.

[31]    Sokol M, Natu V, Kota S, Barsoum MW. On the Chemical Diversity of the MAX Phases. Trends Chem., 2019;1:210-223.

[32]    Kurakevych OO. Superhard phases of simple substances and binary compounds of the B-C-N-O system: from diamond to the latest results (a Review). J Superhard Mater 2009;31:139–157.

[33]    Khazaei M, Arai M, Sasaki T, Estili M, Sakka Y. Trends in electronic structures and structural properties of MAX phases: a first-principles study on $M_2AlC$ (M = Sc, Ti, Cr, Zr, Nb, Mo, Hf, or Ta), $M_2AlN$, and hypothetical $M_2AlB$ phases. J. Phys: Condens. Matter 2014;26:505503.

[34]    Gencer A, Surucu G. Electronic and lattice dynamical properties of $Ti_2SiB$ MAX phase. Mater Res Express 2018;5:076303.

[35]    Surucu G, Investigation of structural, electronic, anisotropic elastic, and lattice dynamical properties of MAX phases borides: An Ab-initio study on hypothetical $M_2AB$ (M = Ti, Zr, Hf; A = Al, Ga, In) compounds. Mater Chem Phys 2018;203:106-117.

[36]    Miao N, Wang J, Gong Y, Wu J, Niu H, Wang S, et al. Computational Prediction of Boron-Based MAX Phases and MXene Derivatives. Chem Mater 2020. DOI: 10.1021/acs.chemmater.0c02139.

[37] Rackl T, Eisenburger L, Niklaus R, Johrendt D. Syntheses and physical properties of the MAX phase boride $Nb_2SB$ and the solid solutions $Nb_2SB_xC_{1-x}$ (x = 0 − 1). Phys Rev Mater 2019;3:054001.

[38] Surucu G, Gencer A, Wang X, Surucu O. Lattice dynamical and thermoelastic properties of $M_2AlB$ (M = V, Nb, Ta) MAX phase borides. J Alloys Compd 2020;819: 153256.

[39] Rackl T, Johrendt D. The MAX phase borides $Zr_2SB$ and $Hf_2SB$. Solid State Sci 2020;106:106316.

[40] Hohenberg P, Kohn W. Inhomogeneous Electron Gas. Phys Rev 1964;136:B864.

[41] Kohn W, Sham LJ. Self-consistent equations including exchange and correlation effects. Phys Rev 1965;140:A1133.

[42] Clark SJ, Segall MD, Pickard CJ, Hasnip PJ, Probert MI, Refson K, et al. First principles methods using CASTEP. Z Kristallogr 2005;220:567-570.

[43] Perdew JP, Burke K, Ernzerhof M. Generalized gradient approximation made simple. Phys Rev Lett 1996;77:3865.

[44] Monkhorst HJ, Pack JD. Special points for Brillouin-zone integrations. Phys Rev B 1976;13:5188.

[45] Fischer TH, Almlöf J. General methods for geometry and wave function optimization. J Phys Chem 1992;96:9768–9774.

[46] Nasir MT, Hadi MA, Naqib SH, Parvin F, Islam AKMA, Roknuzzaman M, et al. , Zirconium metal-based MAX phases $Zr_2AC$ (A = Al, Si, P and S): A first-principles study, Int J Mod Phys B 2014;28:1550022.

[47] Bouhemadou A, Khenata R, Structural, electronic and elastic properties of $M_2SC$ (M = Ti, Zr, Hf) compounds. Phys Lett A 2008;372:6448-6452.

[48] Born M. On the stability of crystal lattices. I. Math Proc Camb Philos Soc 1940;36:160-172.

[49] Sun Z, Music D, Ahuja R, Schneider JM. Theoretical investigation of the bonding and elastic properties of nanolayered ternary nitrides. Phys Rev B 2005;71:193402.

[50] Cover MF, Warschkow O, Bilek MMM, McKenzie DR. A comprehensive survey of $M_2AX$ phase elastic properties. J Phys: Condens Matter 21 (2009) 305403.

[51] Pettifor DG. Theoretical predictions of structure and related properties of intermetallics. J Mater Sci Tech. 1992;8:345-349.

[52] Hill R. The elastic behaviour of a crystalline aggregate. Proc Phys Soc London A 1952;65:349.

[53] Ali MA, Roknuzzaman M, Nasir MT, Naqib SH, Islam AKMA; Structural, elastic, electronic and optical properties of $Cu_3MTe_4$ (M=Nb, Ta) sulvanites: An ab initio study. Int J Mod Phys B 2016;30:1650089.

[54] Bouhemadou A. First-principles study of structural, electronic and elastic properties of $Nb_4AlC_3$. Braz J Phys 2010;40:52-57.

[55] Ali MA, Islam AKMA, Ali MS. Ni-rich nitrides $ANNi_3$ (A = Pt, Ag, Pd) in comparison with superconducting $ZnNNi_3$; J Sci Res 2012;4:1-10.

[56] Ali MA, Islam AKMA, Jahan N, Karimunnesa S. First-principles study of SnO under high pressure. Int J Mod Phys B 2016;30:1650228.

[57] Frantsevich IN, Voronov FF, Bokuta SA. Elastic constants and elastic moduli of metals and insulators handbook, Naukova Dumka, Kiev.1983:60-180.

[58] Pugh SF. XCII. Relations between the elastic moduli and the plastic properties of polycrystalline pure metals. Philos Mag 1954;45:823-843.

[59] Chen XQ, Niu H, Li D, Li Y. Modeling hardness of polycrystalline materials and bulk metallic glasses. Intermetallics 2011;19:1275-1281.

[60] Gao FM. Theoretical model of intrinsic hardness. Phys Rev B 2006;73:132104.

[61] Gou H, Hou L, Zhang J, Gao F. Pressure-induced incompressibility of ReC and effect of metallic bonding on its hardness. Appl Phys Lett 2008;92:241901.

[62] Ali MA, Hadi MA, Hossain MM, Naqib SH, Islam AKMA. Theoretical investigation of structural, elastic and electronic properties of ternary boride MoAlB; Phys Stat Sol B 2017;254:1700010.

[63] Chowdhury A, Ali MA, Hossain MM, Uddin MM, Naqib SH, Islam AKMA. Predicted MAX phase $Sc_2InC$: dynamical stability, vibrational and optical properties; Phys Stat Sol B 2018;255:1700235.

[64] Barua P, Hossain MM, Ali MA, Uddin MM, Naqib SH, Islam AKMA. Effects of transition metals on physical properties of M2BC (M = V, Nb, Mo and Ta): a DFT calculation. J Alloys Compd 2019;770:523-534.

[65] Nasir MT, Hadi MA, Rayhan MA, Ali MA, Hossain MM, Roknuzzaman M et al. First-principles study of superconducting ScRhP and ScIrP pnictides. Phys Stat Sol B 2017;254:1700336.

[66] Sultana F, Uddin MM, Ali MA, Hossain MM, Naqib SH, Islam AKMA. First principles study of $M_2InC$ (M = Zr, Hf and Ta) MAX phases: The effect of M atomic species. Results Phys 2018;11:869-876.

[67] Ali MA, Naqib SH. Recently synthesized $(Ti_{1-x}Mo_x)_2AlC$ ($0 < x < 0.20$) solids solutions: deciphering the structural, electronic, mechanical and thermodynamic properties via ab initio simulations. RSC Adv 2020;10:31535–31546.

[68] Mulliken RS, Electronic population analysis on LCAO-MO molecular wave functions. IV. bonding and antibonding in LCAO and valence-bond theories. J Chem Phys 23;1955:2343.

[69] Ledbetter H, Migliori A. A general elastic-anisotropy measure. J Appl Phys 2006;100:063516.

[70] Chang J, Zhao GP, Zhou XL, Liu K, Lu LY. Structure and mechanical properties of tantalum mononitride under high pressure: A first-principles study. J Appl Phys 2012;112:83519.

[71] Gaillac R, Pullumbi P, Coudert FX. ELATE: an open-source online application for analysis and visualization of elastic tensors. J Phys Condens Matter 2016;28:275201.

[72] Ledbetter HM. Elastic properties of zinc: A compilation and a review. J Phys Chem Ref Data 1977;6:1181.

[73] Islam AKMA, Sikder AS, Islam FN. NbB$_2$: a density functional study. Phys Lett A 2006;350:288-292.

[74] Wang JY, Zhou YC, Liao T, Lin ZJ. First-principles prediction of low shear-strain resistance of Al$_3$BC$_3$: A metal borocarbide containing short linear BC$_2$ units. Appl Phys Letts 2006;89:021917.

[75] Chung DH, Buessem WR. in "Anisotropy in Single Crystal Refractory Compound", Vol-2 edited by F. W. Vahldiek, S. A. Mersol (Plenum Press, New York) 1968:217.

[76] Ranganathan SI, Ostoja-Starzewski M. Universal Elastic Anisotropy Index. Phys Rev Lett 2008;101:055504.

[77] Ali MA, Islam AKMA. Sn$_{1-x}$Bi$_x$O$_2$ and Sn$_{1-x}$Ta$_x$O$_2$ ($0 \leq x \leq 0.75$): A first-principles study. Phys B 2012;407:1020-1026.

[78] Ali MA, Jahan N, Islam AKMA. Sulvanite compounds Cu$_3$TMS$_4$ (TM= V, Nb and Ta): Elastic, electronic, optical and thermal properties using first-principles method. J Sci Res 2014;6:407-419.

[79] Akter K, Parvin F, Hadi MA, Islam AKMA. Insights into the predicted Hf$_2$SN in comparison with the synthesized MAX phase Hf$_2$SC: A comprehensive study. Comput Condens Matter 2020;24:e00485.

[80] Li S, Ahuja R, Barsoum MW, Jena P, Johansson B. Optical properties of Ti$_3$SiC$_2$, Ti$_3$SiC$_2$ and Ti$_4$AlN$_3$, Appl Phys Lett 2008;92:221907.

[81] Ali MA, Ali MS, Uddin MM. First-principles study of elastic, electronic and optical properties of metastable Ti$_5$SiC$_4$. Ind J Pure Appl Phys 2016;54:386.

[82] Togo A, Tanaka I. First principles phonon calculations in materials science. Scrip Mater 2015;108:1-5.

[83] E. I. Isaev, QHA Project, https://qe-forge.org/qha.

[84] Ali MA, Nasir MT, Khatun MR, Islam AKMA, Naqib SH. Ab initio Investigation of Vibrational, Thermodynamic, and Optical properties of Sc$_2$AlC MAX compound. Chinese Phys B 2016;25:103102.

[85] Slack GA, The Thermal Conductivity of Nonmetallic Crystals, in: H. Ehrenreich, F. Seitz, D. Turnbull (Eds.), Solid State Physics, Academic Press, 1979:1–71.

[86] Clarke DR, Materials selection guidelines for low thermal conductivity thermal barrier coatings, Surf Coat Tech 2003;163–164:67–74.

[87] Blanco MA, Francisco E, Luaña V. GIBBS: isothermal-isobaric thermodynamics of solids from energy curves using a quasi-harmonic Debye model, Comput Phys Commun 2004;158:57–72.

[88] Delaire O, May AF, McGuire MA, Porter WD, Lucas MS, Stone MB, et al. Phonon density of states and heat capacity of La$_{3-x}$Te$_4$, Phys Rev B 2009;80:184302.

[89] Plirdpring T, Kurosaki K, Kosuga A, Day T, Firdosy S, Ravi V, et al. Chalcopyrite CuGaTe$_2$: A high-efficiency bulk thermoelectric material, Adv Mater 2012;24:3622-3626.

[90] Xue L, Ren YM, He JR, Xu SL. First-principles investigation of the effects of strain on elastic, thermal, and optical properties of $CuGaTe_2$, Chinese Phys B. 2017;26:067103.

[91] Rayhan MA, Ali MA, Naqib SH, Islam AKMA. First-principles study of Vickers hardness and thermodynamic properties of $T_{i3}Sn_{C2}$ polymorphs; J Sci Res 2015;7:53-64.

[92] Zhou Y, Xiang H, Lu X, Feng Z, Li Z. Theoretical prediction on mechanical and thermal properties of a promising thermal barrier material: $Y_4Al_2O_9$, J Adv Ceram 2015;4:83–93.

[93] Zhan X, Li Z, Liu B, Wang J, Zhou Y, Hu Z. Theoretical Prediction of Elastic Stiffness and Minimum Lattice Thermal Conductivity of $Y_3Al_5O_{12}$, $YAlO_3$ and $Y_4Al_2O_9$. J Am Ceram Soc 2012;95:1429–1434.

[94] Ren Y, Hu Y, Zeng H, Xue L. Prediction on the physical properties of $CuInS_2$ with various anion positions. Curr Appl Phys. 2018;18:304–309.